\documentclass{aa}
\usepackage{graphicx}
\usepackage{txfonts}

\begin{document}
  \title{
    Extrasolar planets and brown dwarfs around  A-F type stars 
    \thanks{Based on observations made with the {\small ELODIE} spectrograph at
      the Observatoire de Haute-Provence
      (CNRS, France) and with the {\small HARPS} spectrograph at La Silla
      Observatory (ESO, Chile) under program ID 073.C-0733.}
  }
  
  \subtitle{I. Performances of radial velocity measurements, first analyses of variations.}
  
  \author{
    F. Galland \inst{1,2}
    \and
    A.-M. Lagrange \inst{1}
    \and
    S. Udry \inst{2}
    \and
    A. Chelli \inst{1}
    \and
    F. Pepe \inst{2}
    \and
    D. Queloz \inst{2}
    \and
    J.-L. Beuzit \inst{1}
    \and
    M. Mayor \inst{2}
  }
  
  \offprints{
    F. Galland,\\
    \email{Franck.Galland@obs.ujf-grenoble.fr}
  }

  \institute{
    Laboratoire d'Astrophysique de l'Observatoire de Grenoble,
    Universit\'e Joseph Fourier, BP 53, 38041 Grenoble, France
    \and
    Observatoire de Gen\`eve, 51 Ch. des Maillettes, 1290 Sauverny, Switzerland
  }
  
  \date{Received 25 February 2005 / Accepted 23 August 2005}

  \abstract{
    We present the performances of a radial velocity measurement method
    that we developed for A-F type stars.
    These perfomances are evaluated through an extensive set of simulations,
    together with actual radial
    velocity observations of such stars using the {\small ELODIE} and {\small HARPS}
    spectrographs. We report the case of stars constant in radial
    velocity, the example of a binary detection on HD\,48097 (an
    A2V star, with $v\sin{i}$ equal to 90 km\,s$^{\rm -1}$) and a confirmation of
    the existence of a 3.9~M$_{\rm Jup}$ planet orbiting around HD\,120136 (Tau
    Boo). The instability strip problem is also discussed.
    We show that with this method, it is in principle possible to
    detect planets and brown dwarfs around A-F type stars, thus
    allowing further study of the
    impact of stellar masses on planetary system formation over a wider range of stellar
    masses than is currently done.

  }
  
  \maketitle
  
  \section{Introduction}

  Since the discovery of the first exoplanet around a solar-like star
  a decade ago (\cite{MayorQ95}), more than 150 planets have been
  found by radial velocity surveys \footnote{A list of
  discovered planets updated by Jean Schneider is available  at
  http://www.obspm.fr/encycl/cat1.html}. These surveys
  focus on late type stars ($\ga$ F8) as these stars exhibit numerous
  lines with low rotational broadening. 
  General characteristics of planet masses, distances to star,
  eccentricities (see e.g. \cite{Udry03}; \cite{Marcy03}), as well as
  characteristics of the stars hosting those giant planets
  (e.g. metallicity; \cite{Santos03}) were derived and allow
  theoreticians to constrain planetary system
  formation and evolution (e.g. planet migration) around solar type stars.

  A general and fundamental question concerning planet formation is the
  impact of the mass of the central star on the formation and evolution
  process. We know that the disks around these
  different types of stars do not have the same
  properties at similar ages: TTauri disks of a few
  Myr appear to be less evolved than those around massive stars
  also of a few Myr such as HD\,141569 or HR\,4796; this tends to show that
  these disks, as the parent stars, evolve more rapidly
  (\cite{LA04}). The occurrence and time scale of planet
  formation have to be investigated and compared. 

\begin{figure*}[t!]
    \centering
    \includegraphics[width=0.32\hsize]{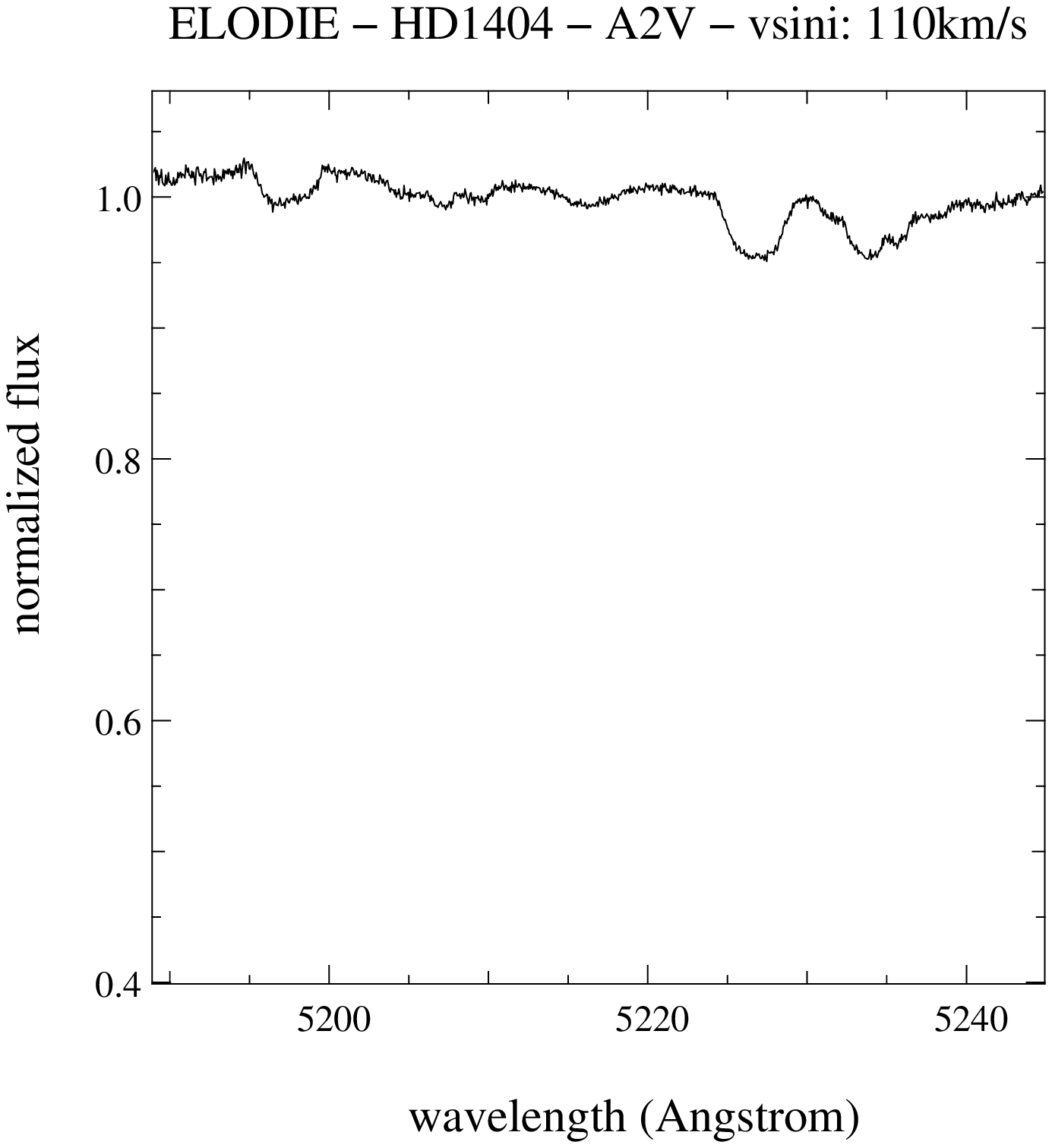}
    \includegraphics[width=0.32\hsize]{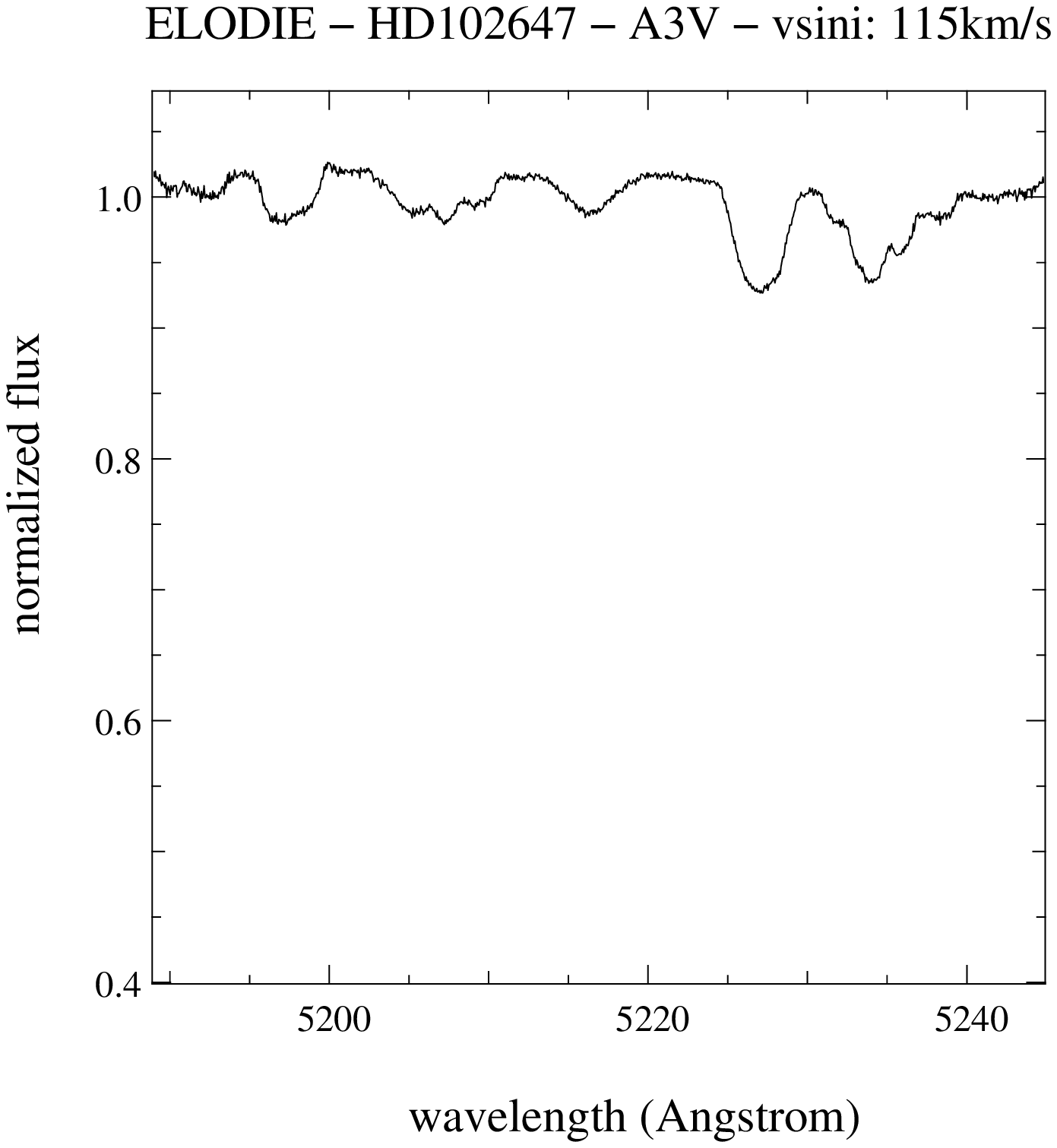} 
    \includegraphics[width=0.32\hsize]{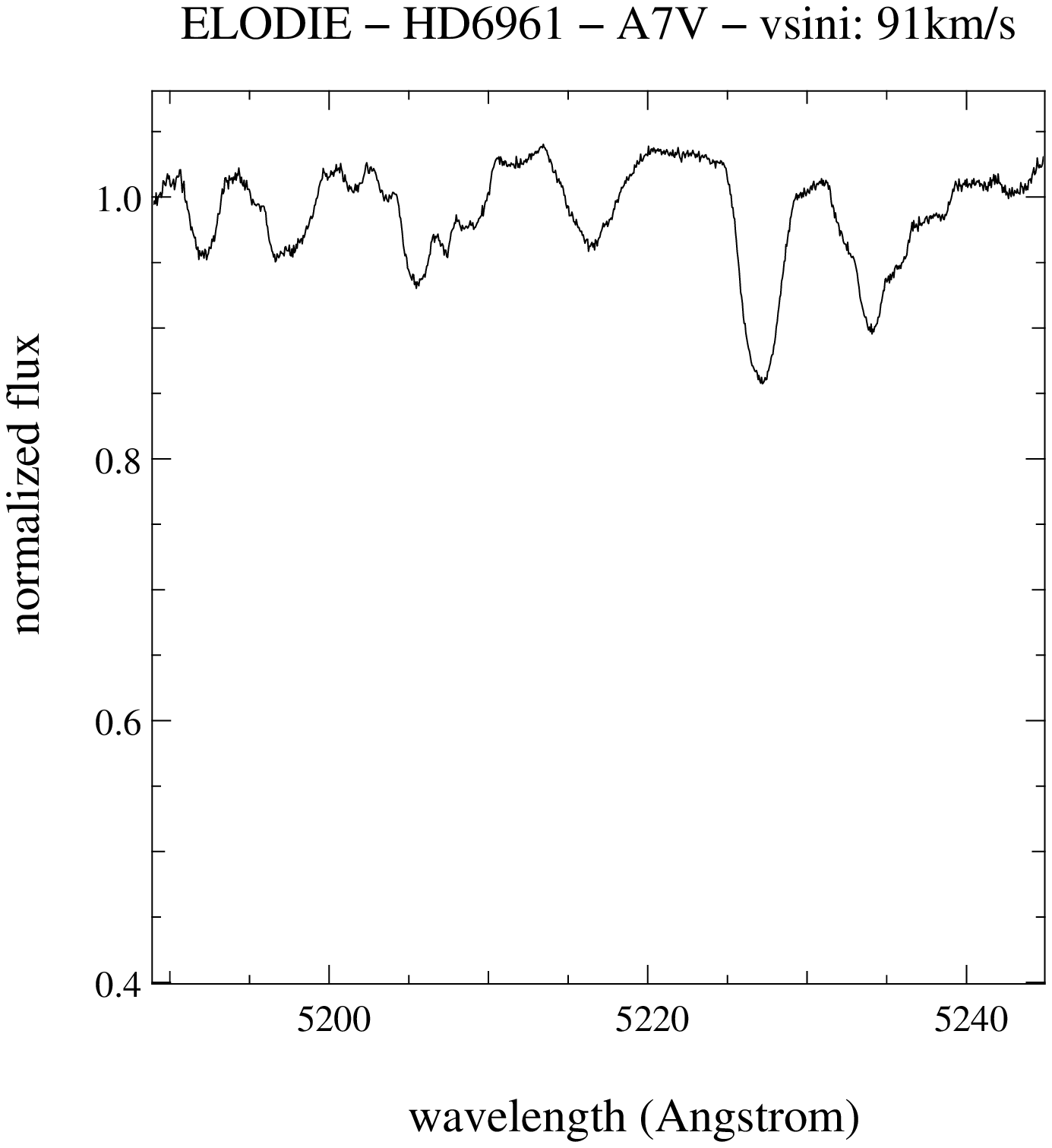}

    \vspace{0.2cm}

    \includegraphics[width=0.32\hsize]{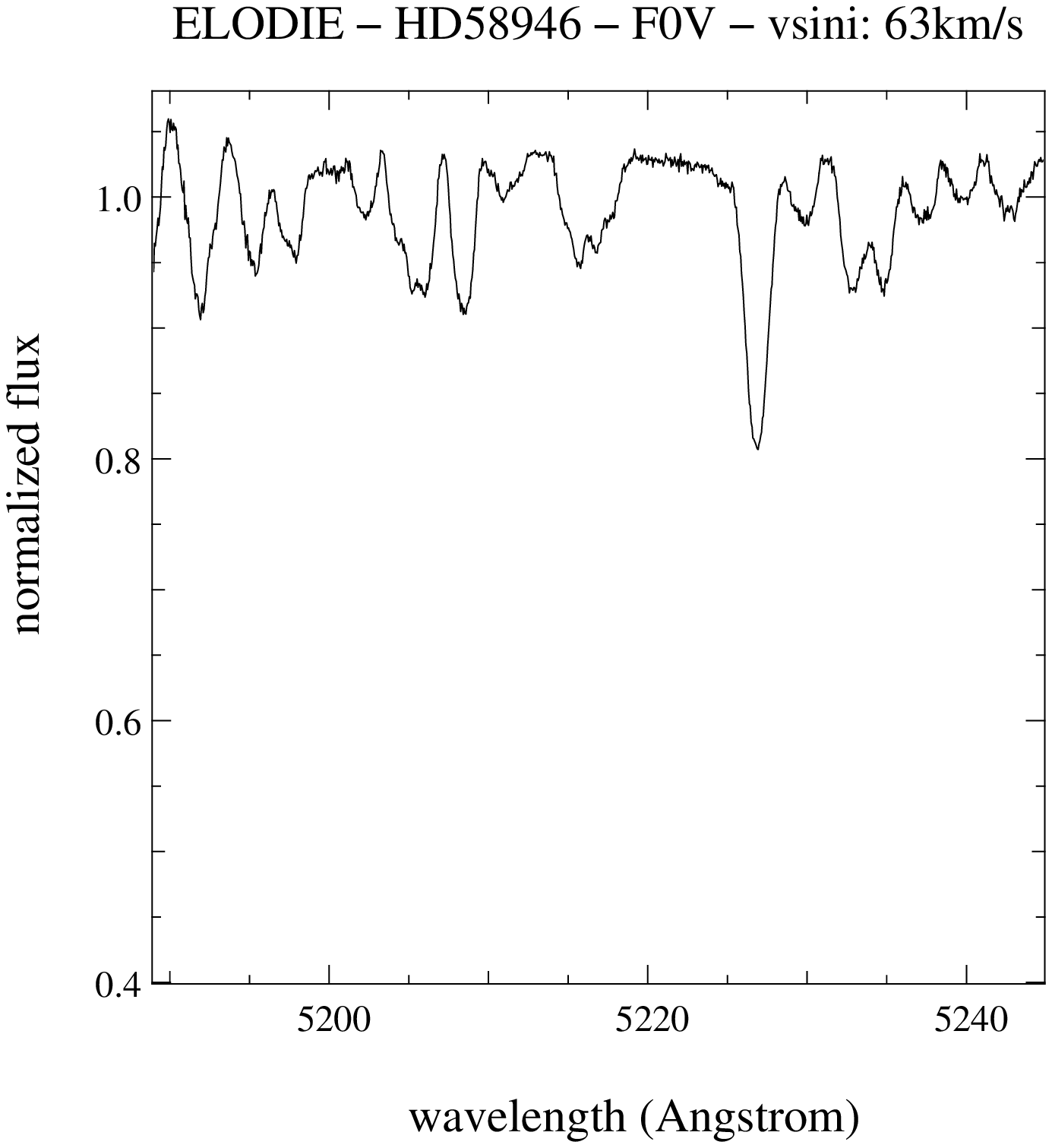} 
    \includegraphics[width=0.32\hsize]{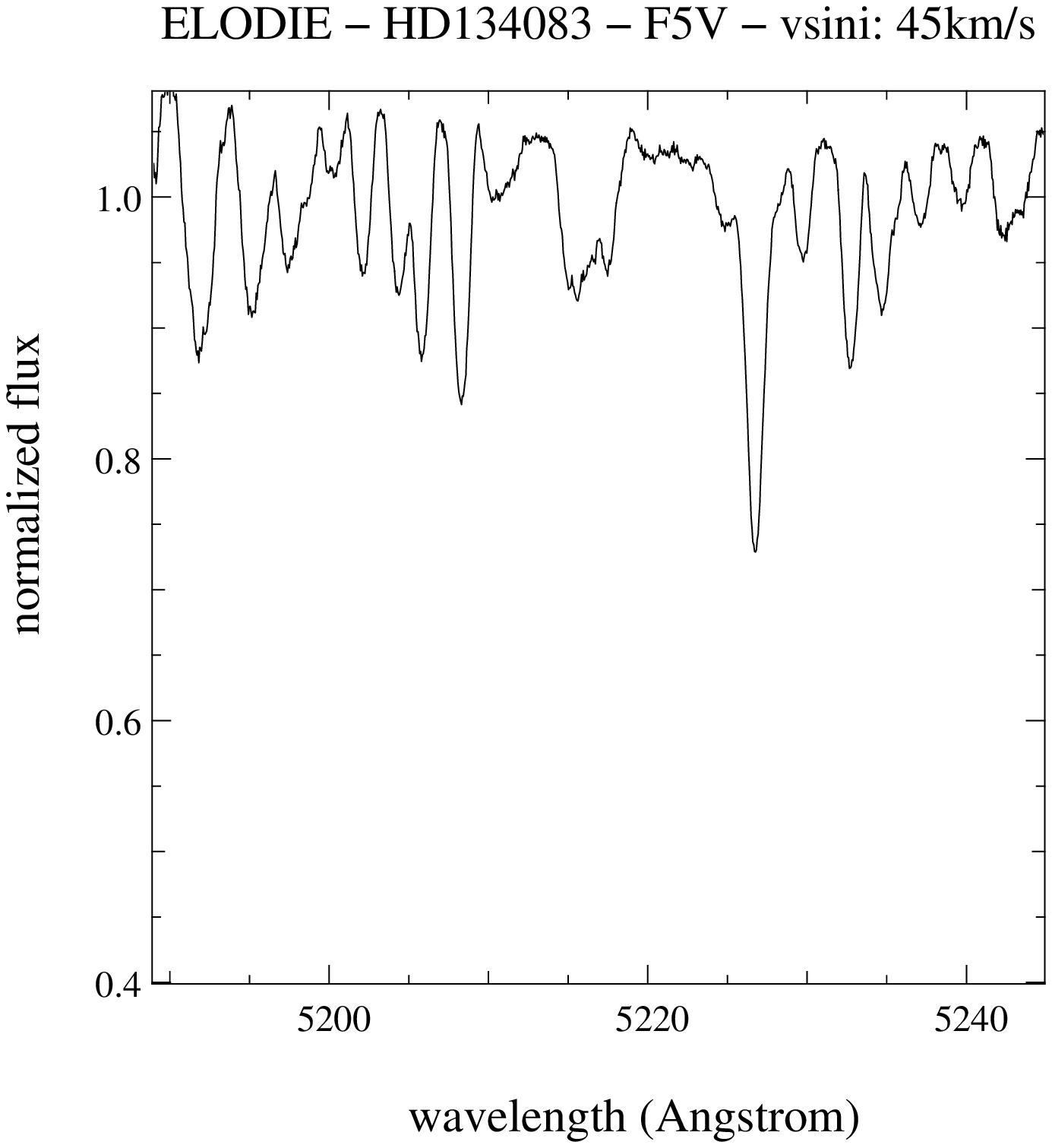} 
    \includegraphics[width=0.32\hsize]{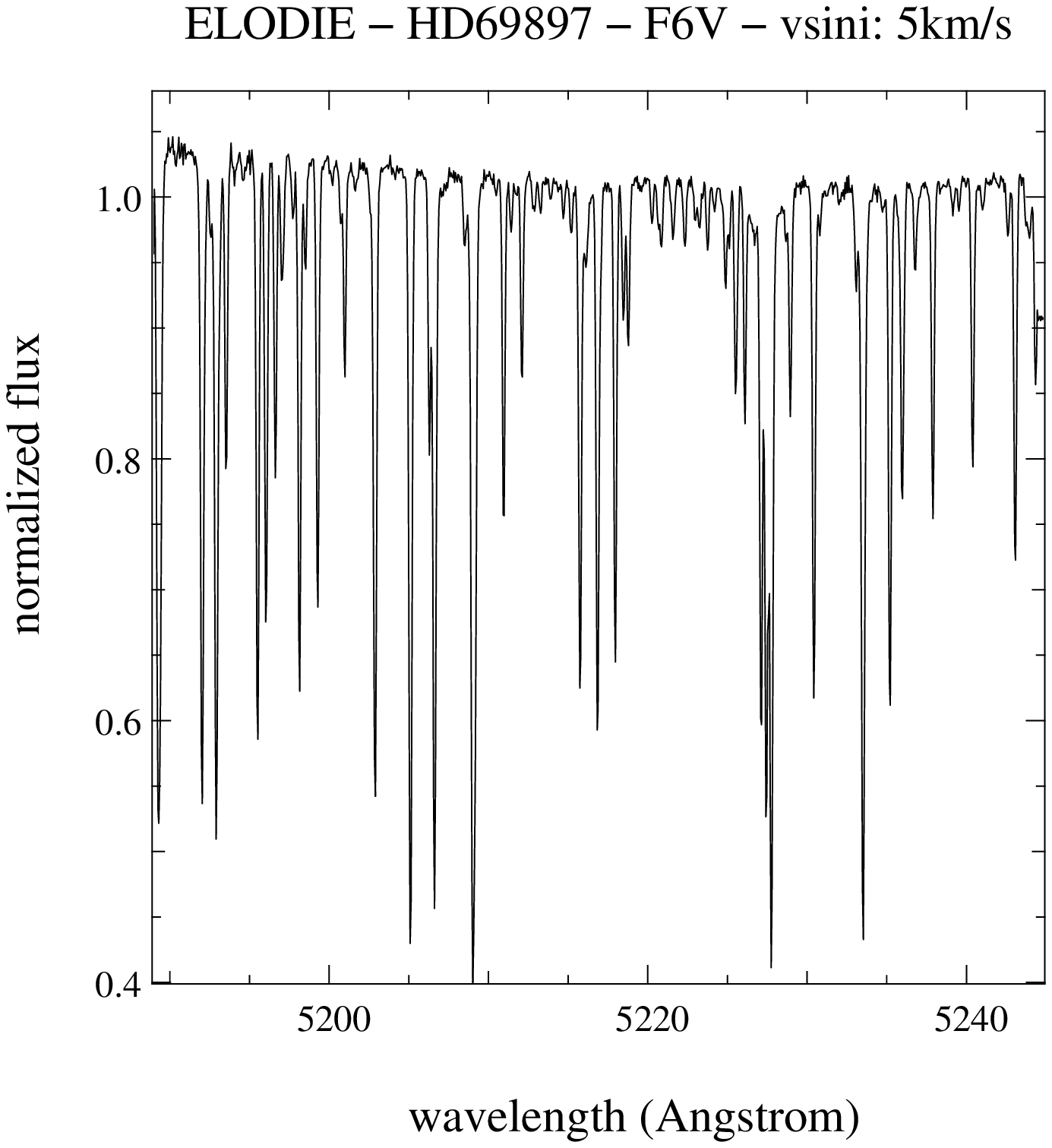}

    \caption{Examples of spectra acquired with {\small ELODIE}. Note the effect
    of the spectral type and vsini on the number, depth and width of the lines.}
    \label{sp_E_o40}
  \end{figure*}

  Looking for planets around early type stars is a difficult task. So
  far, the studies have been limited to giant stars (\cite{Sato03},
  \cite{Lovis04}). These stars have small
  rotational velocities, but a large radius resulting in minimum
  possible orbital periods of the order of 100 days or slightly less. In
  a complementary way, to focus on main sequence
  stars allows us to access smaller orbital periods and to address the question
  of evolutionary time scale.
  These high-mass main sequence stars have not been investigated so far as
  they exhibit fewer lines that are
  generally broadened by high rotational velocities
  (typically 100 - 200 km\,s$^{\rm -1}$ for A-type stars; see
  Fig.~\ref{sp_E_o40}). It was then thought that the
  radial velocity method could not be
  applied to those objects. Indeed, the method to process the data and
  extract the Doppler information for low-mass stars with the
  cross-correlation method, is not
  straightforwardly applicable to more massive stars
  (\cite{Griffin00}).

  A new method for radial velocity measurements was introduced a few
  years ago (\cite{Chelli00}). It consists of correlating, in Fourier space,
  each spectrum of the target star and a reference spectrum specific to that star (built
  e.g. by summing all the available spectra of the star). This
  method had been applied to solar-type stars; we have adapted it for
  earlier type stars (Section~2). 
  We then performed radial velocity observations of A-F type stars,
  using the fiber-fed echelle spectrographs
  {\small ELODIE} (\cite{Baranne96}), mounted on the 1.93-m telescope at the
  Observatoire de Haute Provence (CNRS, France) in the northern
  hemisphere, and {\small HARPS} (\cite{Pepe02}), recently installed on the 3.6-m ESO
  telescope at La Silla Observatory (ESO, Chile) in the southern
  hemisphere. 

  We present here the performances obtained by applying this
  method to these spectroscopic observations of A-F main sequence stars.
  We demonstrate in particular
  that it should be possible to detect planets and
  brown dwarfs around A-F type stars. Results obtained with
  simulations are provided in Section~3. We confirm in Section~4 the accuracy of
  the computed radial velocities and corresponding
  uncertainties in real cases, with examples of stars constant in radial
  velocity, the case of a binary detection and the  confirmation of a
  3.9~M$_{\rm Jup}$ planet orbiting around HD\,120136 (Tau Boo).
  Furthermore, we discuss the first trends observed in radial
  velocity variations as a function of the spectral type in the range
  [A0-F7]. Finally, we present
  the uncertainties achieved for all stars already observed
  (Section~5), and the corresponding mass detection limits in the frame of
  planet and brown dwarf searches (Section~6).

  \section{Principle of the method}

  To compute the radial velocity, we use the method described in Chelli (2000).   
  Considering a reference spectrum $S_r(\lambda)$ and a Doppler shifted
  one, $S(\lambda)=S_r(\lambda-\lambda{U \over c})$, where $U$ is the
  radial velocity associated with $S(\lambda)$, we derive the cross spectrum:
  $$\widehat{I} (\nu) = \widehat{S}_r(\nu)\widehat{S}^*(\nu) = {\rm e}^{2i\pi\nu\lambda_0{U \over
      c}}|\widehat{S}_{\rm r}(\nu)|^2$$
  The radial velocity is contained in the phase of $\widehat{I}(\nu)$.
  We then look for the velocity $V$ which minimizes in a least square sense the
  imaginary part of the quantity:
  $ \widehat{C}(\nu) = {\rm e}^{-2i\pi\nu\lambda_0{V \over c}}
      \widehat{I}(\nu) $.
  The quantity to be minimized becomes:                               
  $$\chi^2 = \sum_{j}^{} {{Im}^2[\widehat{C}(\nu_j)] \over \sigma^2(\nu_j)} = 
  \sum_{j}^{} {{ sin^2(2i\pi\nu_j\lambda_0{{U-V} \over
  c})|\widehat{S}_{r}(\nu)|^4 } \over \sigma^2(\nu_j)}$$
  which is minimum when V reaches the radial velocity shift U. $\nu_j$
  are discrete frequencies. The photonic radial velocity uncertainty ($\epsilon_{\mathrm{RV-Ph}}$)
  is simultaneously calculated using the photon noise statistic. See
  Chelli (2000) for further details.

   \begin{figure*}[t!]
    \centering
    \includegraphics[width=0.34\hsize]{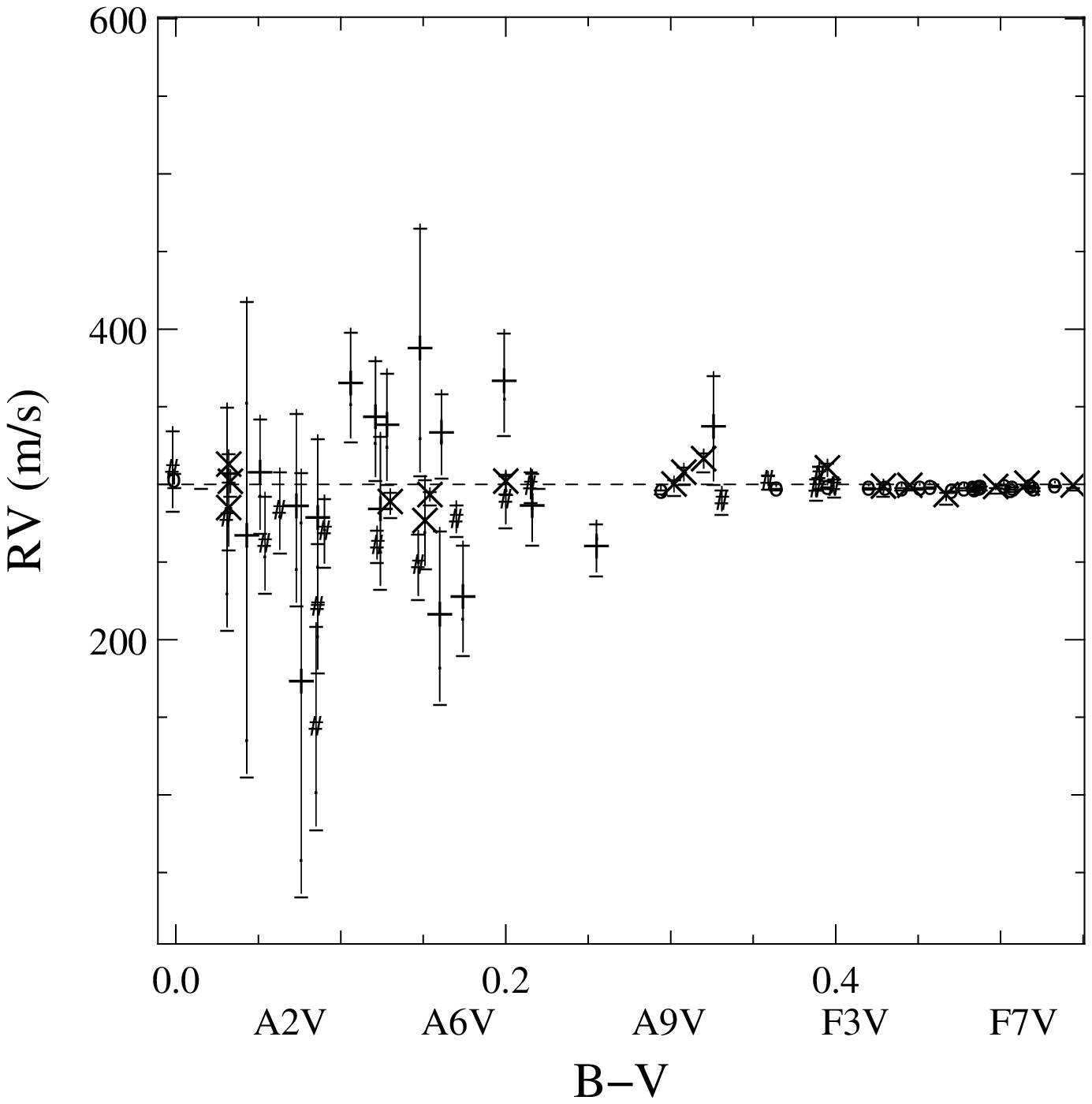}
    \includegraphics[bb=96 271 474 671,width=0.34\hsize,clip]{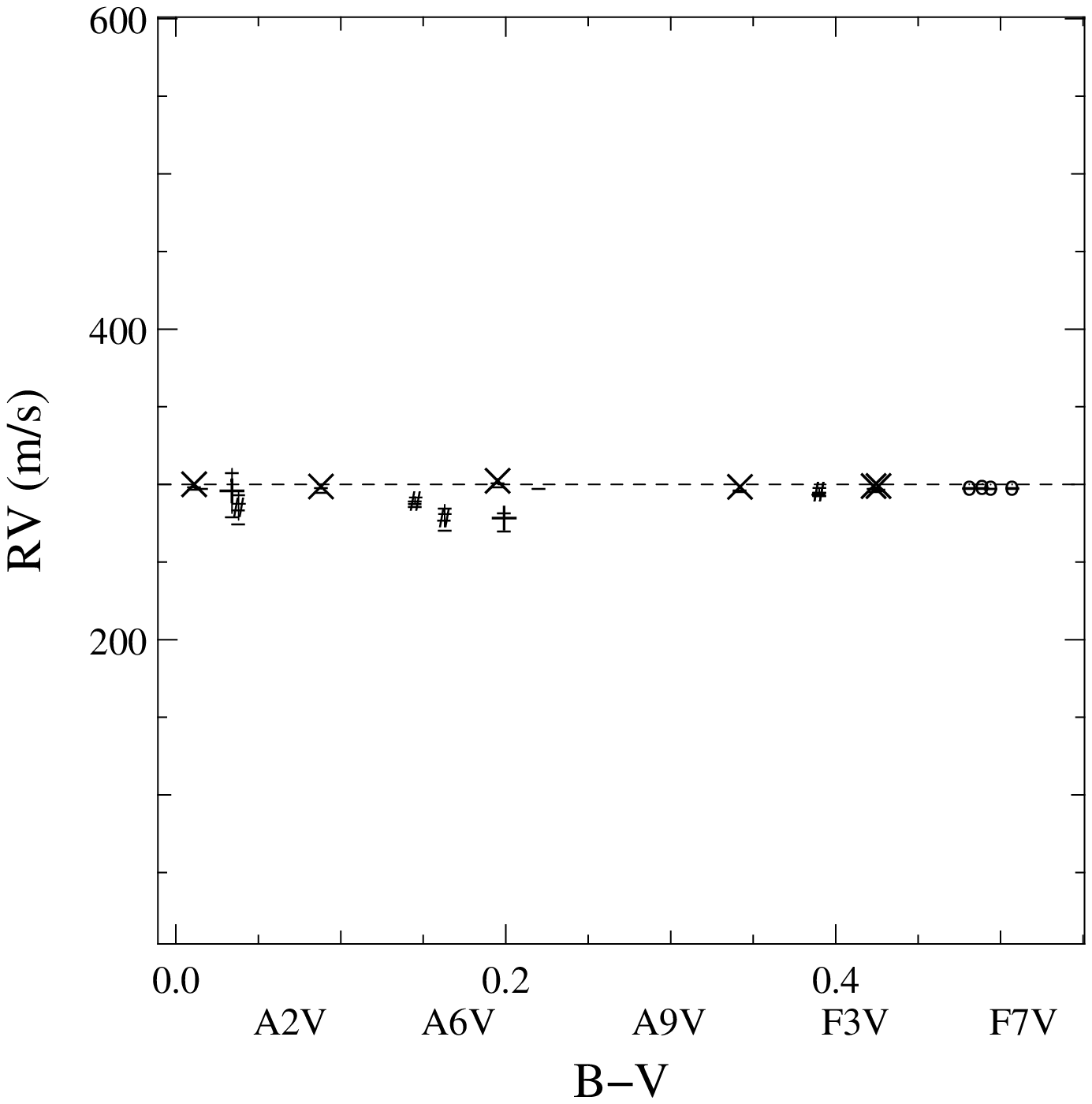} 
    \includegraphics[width=0.34\hsize]{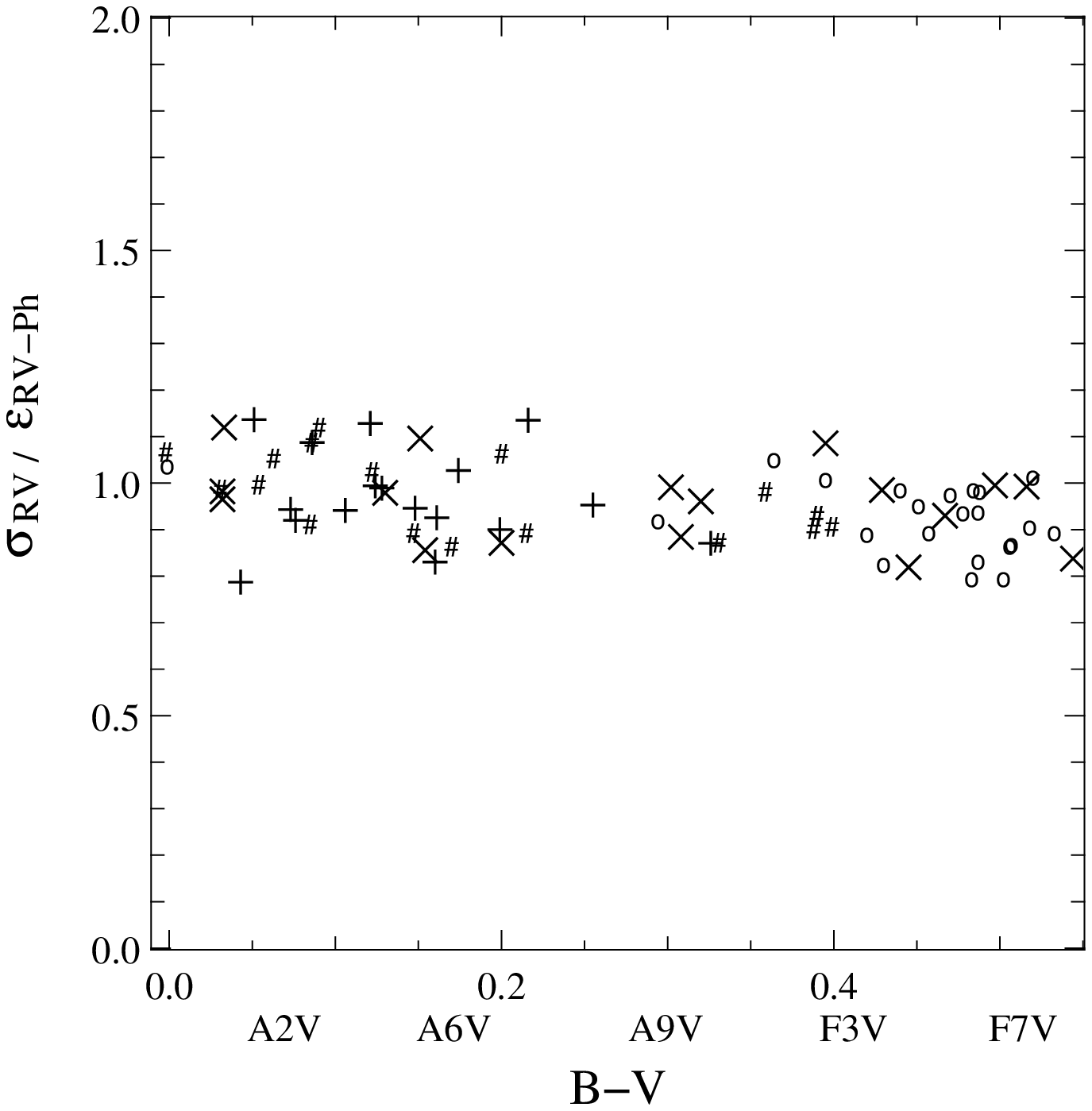}
    \includegraphics[bb=96 271 474 671,width=0.34\hsize,clip]{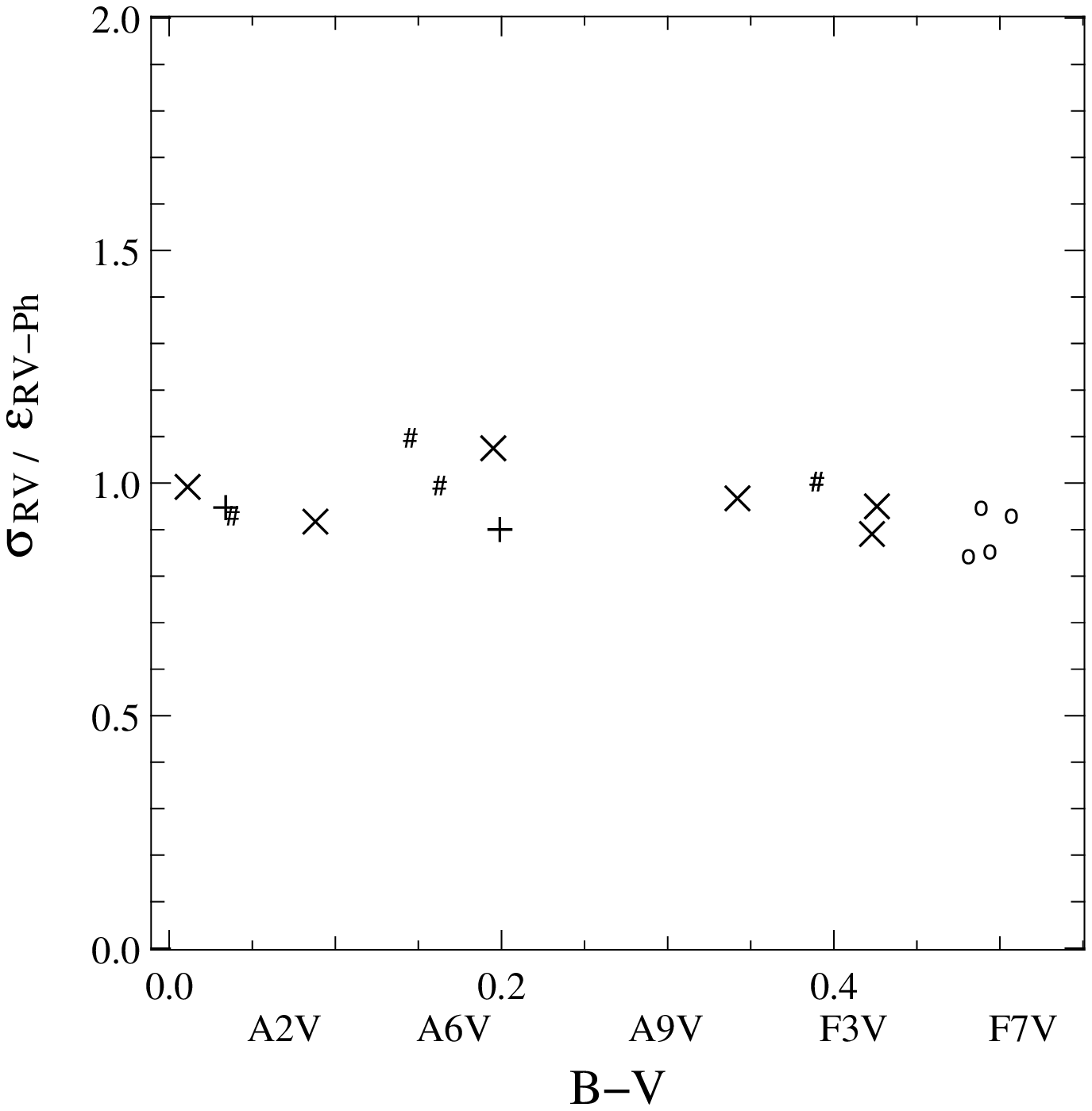} 
    \caption{
      Top: Average computed radial velocity obtained by simulations on
      spectra acquired with {\small ELODIE} (left) and {\small HARPS} (right). Error
      bars from the average of 100 test spectra.
      Bottom: radial velocity dispersions divided by uncertainties obtained
      from simulations on {\small ELODIE} (left) and
      {\small HARPS} (right) spectra.
      Conventions for symbols: $v\sin{i}$ $\leq$
      20~km\,s$^{\rm -1}$~$(o)$, 20~km\,s$^{\rm -1}$~$\leq$ $v\sin{i}$~$\leq$~70~km\,s$^{\rm -1}$
      $(\times)$, 70~km\,s$^{\rm -1}$~$\leq$~$v\sin{i}$~$\leq$~130~km\,s$^{\rm -1}$~$(\#)$,
      130~km\,s$^{\rm -1}$~$\leq$~$v\sin{i}$~$(+)$.
    }
    \label{t_EH_VR_ectpeps}
  \end{figure*}

  Working in the Fourier space allows us to apply frequency cuts,
  reducing the impact of the noise (high frequencies) and of the
  variations of the continuum (low frequencies) due to stellar
  phenomena or instrumental effects. This is particularly interesting in the case
  of A-F type stars.

  \section{Radial velocity measurements and uncertainties estimates: simulations}

  We first present simulations performed to test our radial velocity
  and photon noise uncertainty ($\epsilon_{\mathrm{RV-Ph}}$) determination on individual
  measurements.

  \subsection{Test spectra}

  We consider here the case of identical spectra shifted with a
  constant radial
  velocity. We first consider a spectrum of a given star
  obtained with {\small ELODIE} or {\small HARPS};
  this spectrum is smoothed then duplicated into
  several spectra (here, 101). We add noise to one of these
  spectra at a level corresponding to
  a typical reference spectrum, namely corresponding to the sum of
  100 spectra with S/N equal to 200 in the
  case of {\small ELODIE}, and 400 in the case of {\small HARPS}. We add noise to the
  other 100 spectra at a level corresponding to a typical measurement
  i.e. S/N = 200 with {\small ELODIE} and 400 with {\small HARPS}. Then,
  these spectra are shifted in radial velocity with a given value
  (here, 300 m\,s$^{\rm -1}$), typically induced by the presence of a planet.

  Our method is then used to measure the radial velocities and the corresponding
  uncertainties. For the considered star, we obtain a distribution of
  computed radial
  velocities, characterized with its average and dispersion (if assumed gaussian).  
  The advantages of these tests are: the original
  spectrum corresponds to a real case in terms of
  number and depth of spectral lines, taking into account rotational
  broadening and spectral type which are key parameters;
  the average radial velocity measured can be compared to the shift applied,
  in order to test the radial velocity measurement process; no
  effect other than velocity shift changes the spectra.

  \begin{figure*}[t!]  
    \centering
    \includegraphics[bb= 88 271 484 666,width=0.3\hsize]{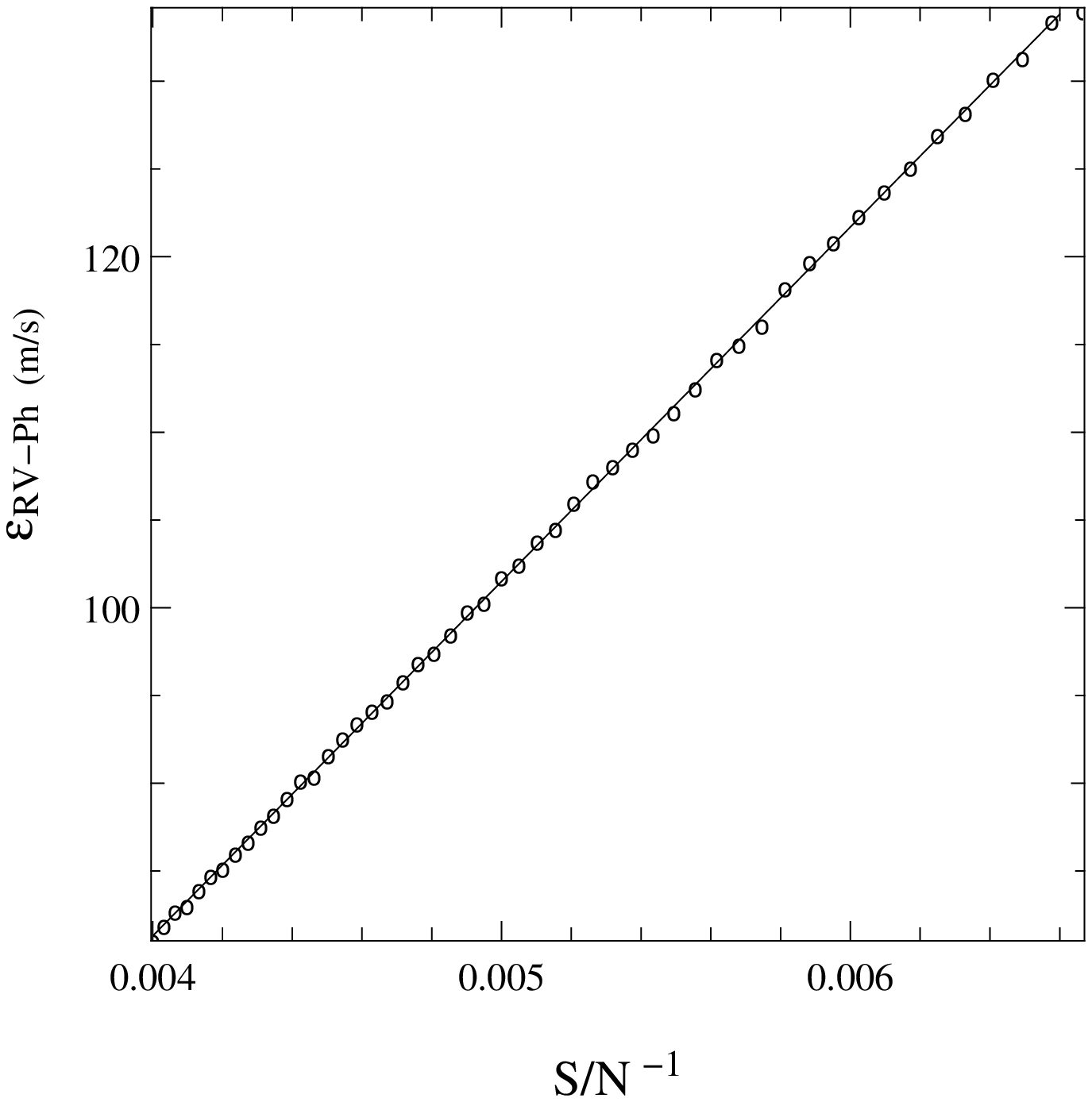}
    \includegraphics[width=0.3\hsize]{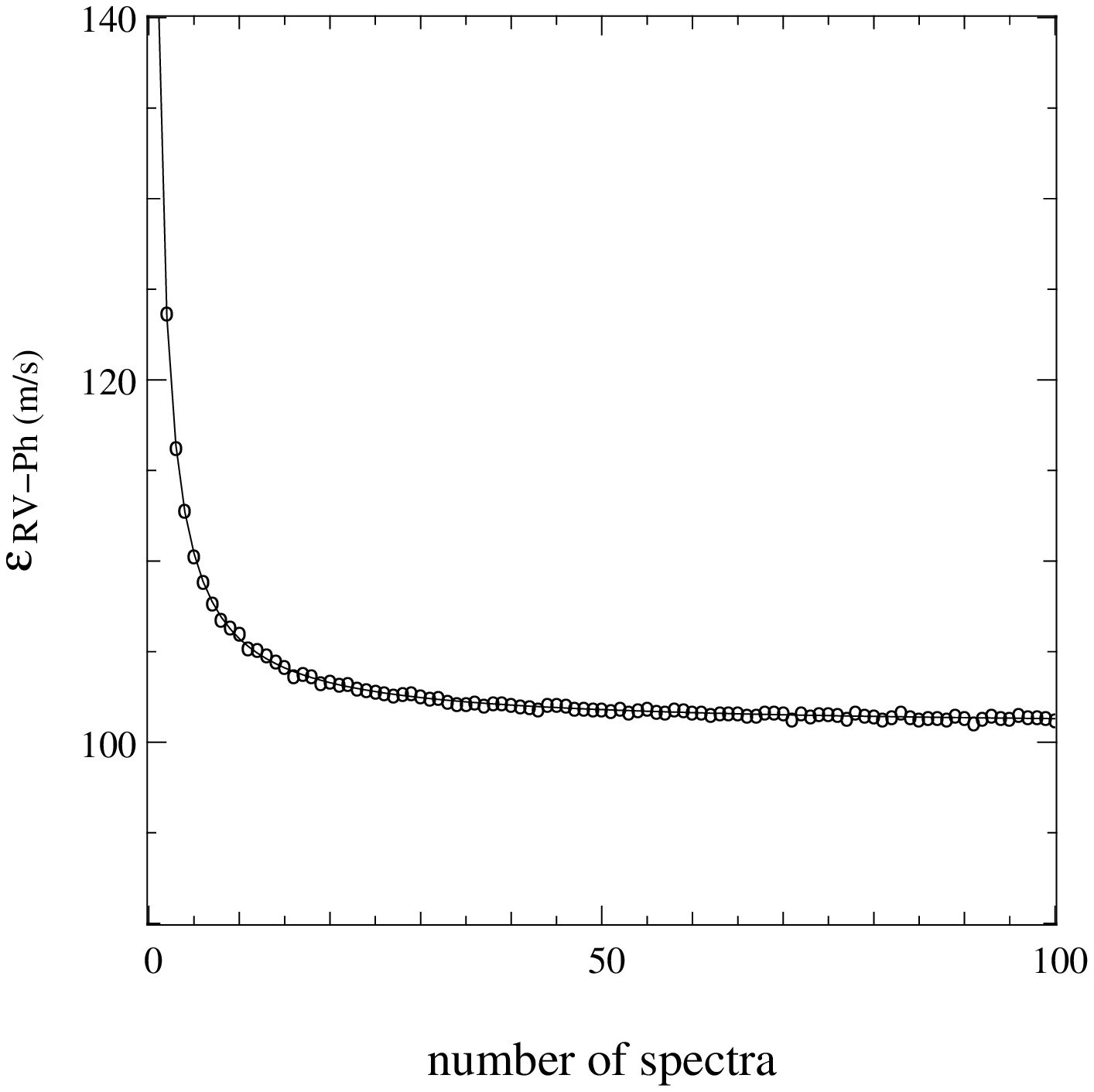}
    \includegraphics[width=0.3\hsize]{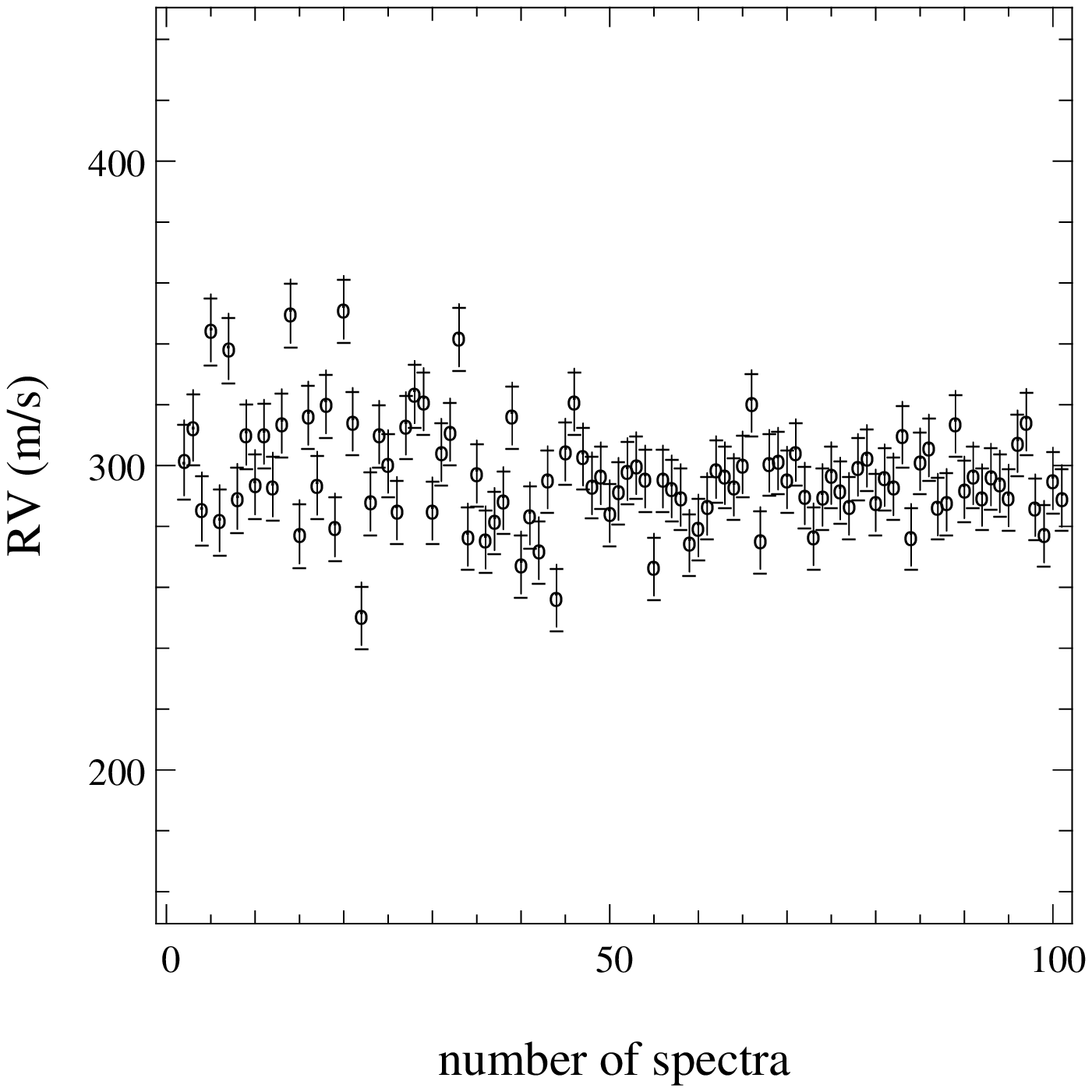}
    \caption{Simulated radial velocity uncertainties versus S/N (left),
    versus the number of spectra used to
    build the reference spectrum (center), and radial velocities
    obtained as a function of this number of spectra (right), in the
    case of an A7V star, $v\sin{i}$ = 90 km\,s$^{\rm -1}$, with {\small ELODIE}.}
    \label{t_epsVR_SB_n}
  \end{figure*}

  \begin{figure*}[t!]
    \centering
    \includegraphics[width=0.35\hsize]{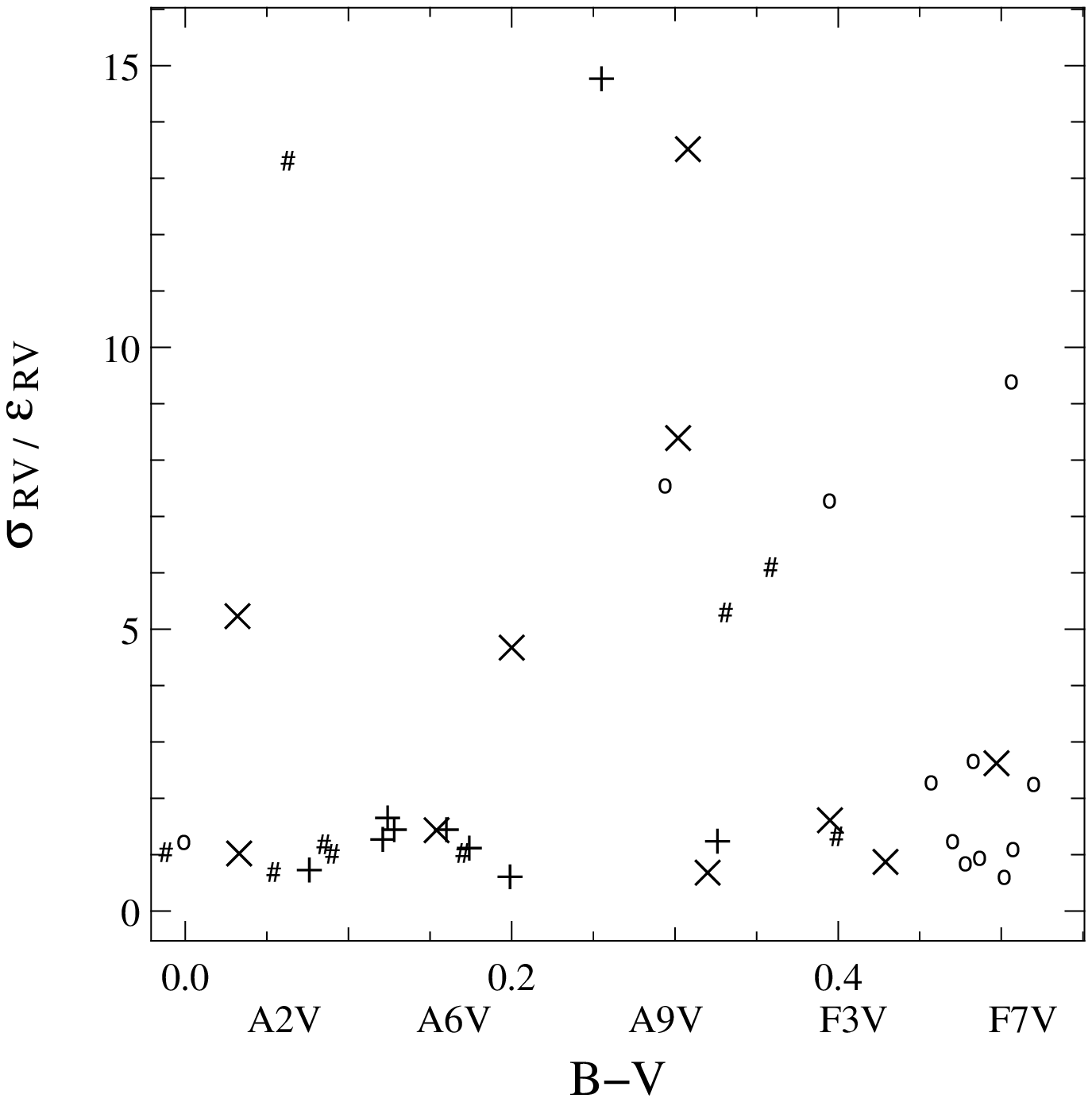}
    \includegraphics[bb=96 271 474 671,width=0.35\hsize,clip]{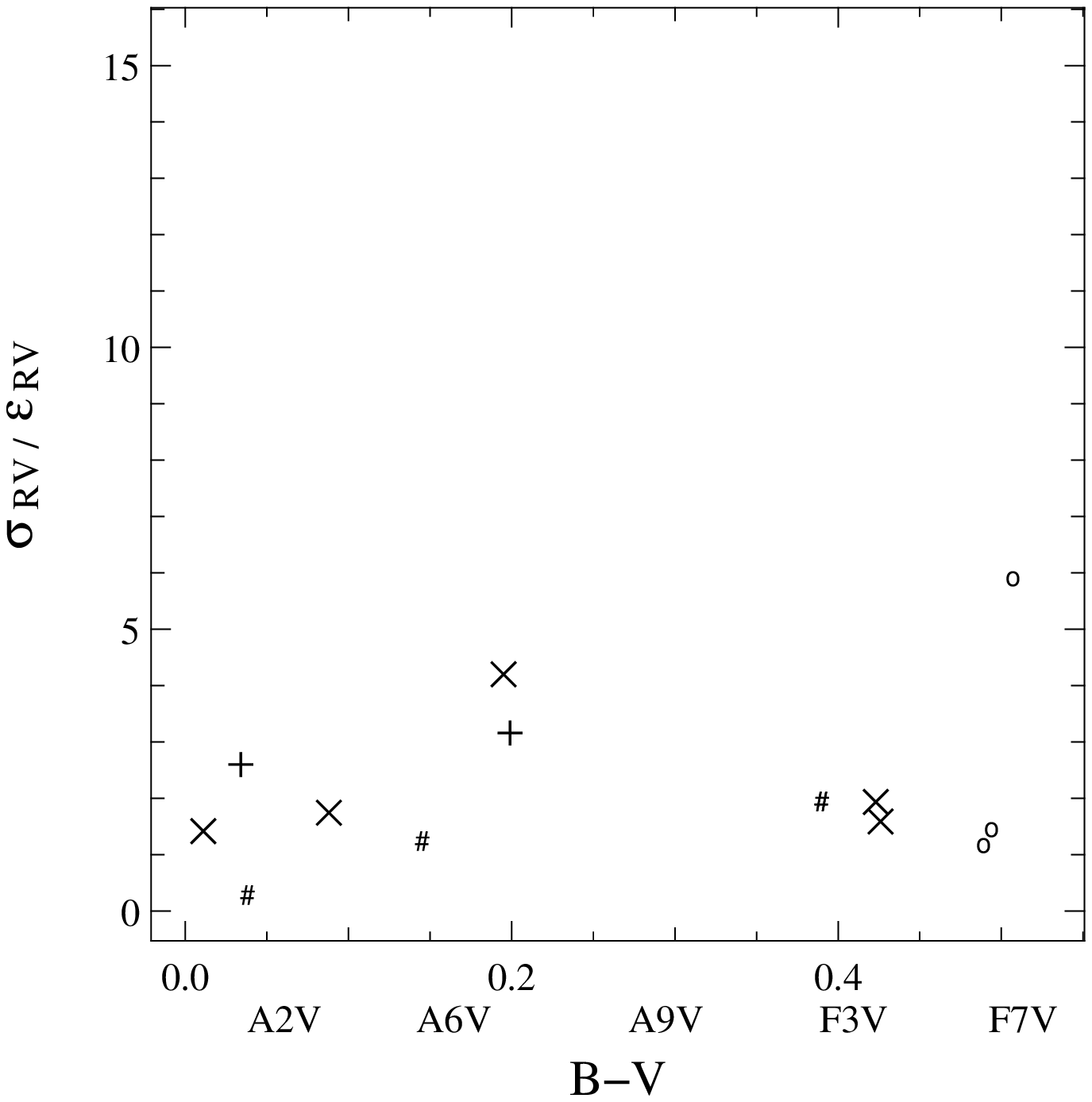}
    \caption{Measured dispersions divided by uncertainties obtained with
      {\small ELODIE} (left) and {\small HARPS} (right), with at least 5
      measurements. S/N are those obtained during observations.
      Symbols are the same as in Fig.~\ref{t_EH_VR_ectpeps}.
    }
    \label{EH_std}
  \end{figure*}

  \subsection{Results}

  The radial velocity measurements obtained are shown in
  Fig.~\ref{t_EH_VR_ectpeps} (top), for different spectral types (ranging between A0V
  and F7V), and different rotational
  velocities ($v\sin{i}$).
  The measurements correspond to the average on the distribution of
  computed radial velocities for a given star. Error bars correspond to
  $\frac{\epsilon_{\mathrm{RV-Ph}}}{\sqrt{100}} =
  \epsilon'_{\mathrm{RV-Ph}}$, as they correspond
  to the 100 test spectra for
  a given star. The average computed radial velocity is consistent with the initial
  shift RV$_\mathrm{o}$ equal to 300 m\,s$^{\rm -1}$, given the error bars.
  This demonstrates the accuracy of the
  computed radial velocities (no systematic error), in the case of identical spectra only
  shifted in radial velocity.

  The associated uncertainties are shown in
  Fig.~\ref{t_EH_VR_ectpeps}. In all cases,
  the radial velocity  dispersion
  $\sigma_{\mathrm{RV}}$ on the radial velocity distribution obtained for each star
  is consistent with  the computed uncertainties
  $\epsilon_{\mathrm{RV-Ph}}$: the distribution of
  $\frac{\sigma_{\mathrm{RV}}}{\epsilon_{\mathrm{RV-Ph}}}$ on the
  results obtained for the different stars is centered on 0.95
  $\pm$ 0.09 for {\small ELODIE}, and on 0.94 $\pm$ 0.1 for {\small HARPS}. This
  demonstrates the quality of the photon noise uncertainty estimates.

  The same kind of tests as above (creation of 100 test spectra shifted
  by 300 m\,s$^{\rm -1}$, and a
  reference spectrum) have been performed to check the dependence of
  $\epsilon_{\mathrm{RV-Ph}}$ on the number n of spectra used to
  compute the reference spectrum (with  $\mathrm{S/N}$ fixed to 200)
  and on $\mathrm{S/N}$ (with n fixed to 100).
  We find that the radial velocity uncertainties behave 
  following these relations:
  $ \epsilon_{\mathrm{RV-Ph}}
  \propto \frac{1}{\mathrm{S/N}}$ where S/N is the signal--to--noise
  per pixel for one spectrum. This is useful when planning
  measurements, regarding the exposure time necessary to reach a
  given $\epsilon_{\mathrm{RV-Ph}}$;
  $ \epsilon_{\mathrm{RV-Ph}}
  \propto \sqrt{\frac{n+1}{n}}$ is nearly constant for $n \geq 10$.

  The results of the computation are displayed in
  Fig.~\ref{t_epsVR_SB_n} as dots: they are in good agreement with
  fits corresponding to these relations. 
  Moreover, the radial velocities are not affected by systematic
  errors, even if the number of spectra used to build the reference
  spectrum is small (Fig.~\ref{t_epsVR_SB_n}).

  \section{Radial velocity dispersions and uncertainties: real case}

  Our aim is to confirm the accuracy of
  the computed radial velocities and corresponding
  uncertainties in real cases.
  We use here the data available so far on our sample of stars
  surveyed with {\small ELODIE} and {\small HARPS}.
  By January 2005, 45 A-F type stars were
  observed at least 5 times with {\small ELODIE} ($\mathrm{S/N}$ equal
  to 196 on average) and 13 with {\small HARPS} ($\mathrm{S/N}$ equal
  to 302 on average). 

  The dispersions obtained, compared to
  the observed radial velocity
  uncertainties, are displayed in Fig.~\ref{EH_std}.
  They are consistent with an accurate computation of the radial velocity
  uncertainties in real cases, because they verify $\sigma_{\mathrm{RV}} \ga
  \epsilon_{\mathrm{RV}}$ in most cases, where
  $\epsilon_{\mathrm{RV}} = \sqrt{\epsilon^2_{\mathrm{RV-Ph}} +
  \epsilon^2_{\mathrm{RV-Ins}}} \;$. In addition to the photon noise,
  $\epsilon_{\mathrm{RV}}$ takes
  into account instrumental (in)stability
  ($\epsilon_{\mathrm{RV-Ins}}$: 6.5 m\,s$^{\rm -1}$ with {\small
  ELODIE}, 1 m\,s$^{\rm -1}$ with {\small HARPS}).

  \begin{table*}[t!]
    \caption[]{Radial velocity (RV) dispersion and
      uncertainties (measured and normalized to the same S/N, 200) for
      a set of stars
      observed with {\small ELODIE} and constant in radial velocity. The
      normalization allows us to focus the comparison between stars
      only on spectral type and $v\sin{i}$.}
    \label{Table_ectp_sigv}
    \begin{center}
      \begin{tabular}{ l c c c c c c }
	\hline
	\hline
	star & spectral & v.sin(i)  & number of  & dispersion on &
	measured RV  & normalized RV      \\
        & type     &   km\,s$^{\rm -1}$       & measurements   &
	measured RV ($m\,s^{\rm -1}$)
	& uncertainties ($m\,s^{\rm -1}$) & uncertainties ($m\,s^{\rm -1}$)\\
	\hline
	HD1404   &  A2V &     110 &       15 &     203 &  280   &     318 \\
	HD102647 &  A3V &     115 &       17 &     137 &  132   &     222 \\
	HD6961   &  A7V &      91 &       15 &      88 &   83   &     104 \\
	HD58946  &  F0V &      63 &        6 &      32 &   47   &      66 \\
	HD134083 &  F5V &      45 &       11 &      34 &   39   &      44 \\
	HD69897  &  F6V &       5 &        9 &     7.7 &  7.7   &     7.4 \\
	\hline
	\hline
      \end{tabular}
    \end{center}
  \end{table*}
   
  \begin{figure*}[t!]
    \centering
    \includegraphics[width=0.32\hsize]{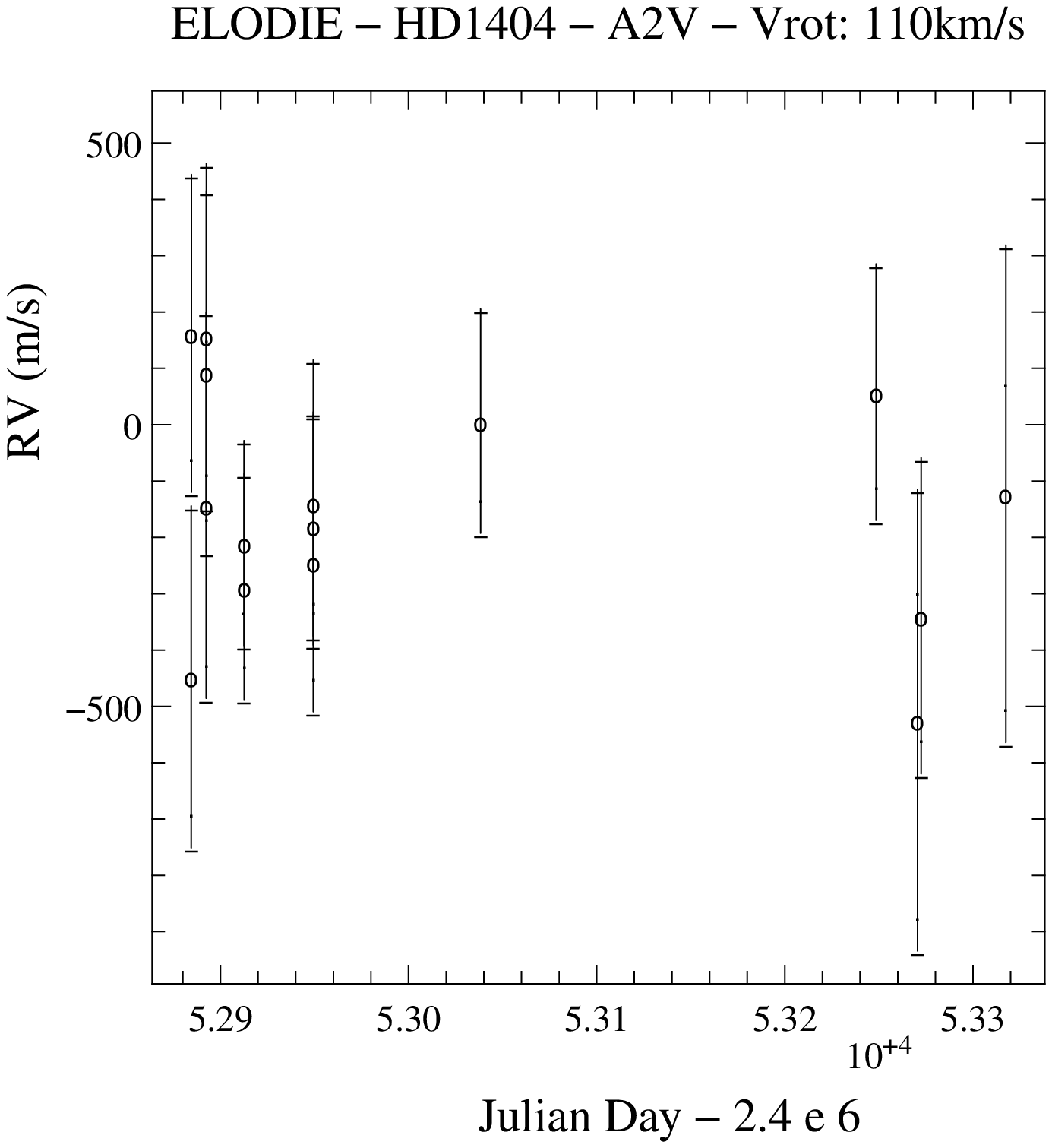}
    \includegraphics[width=0.32\hsize]{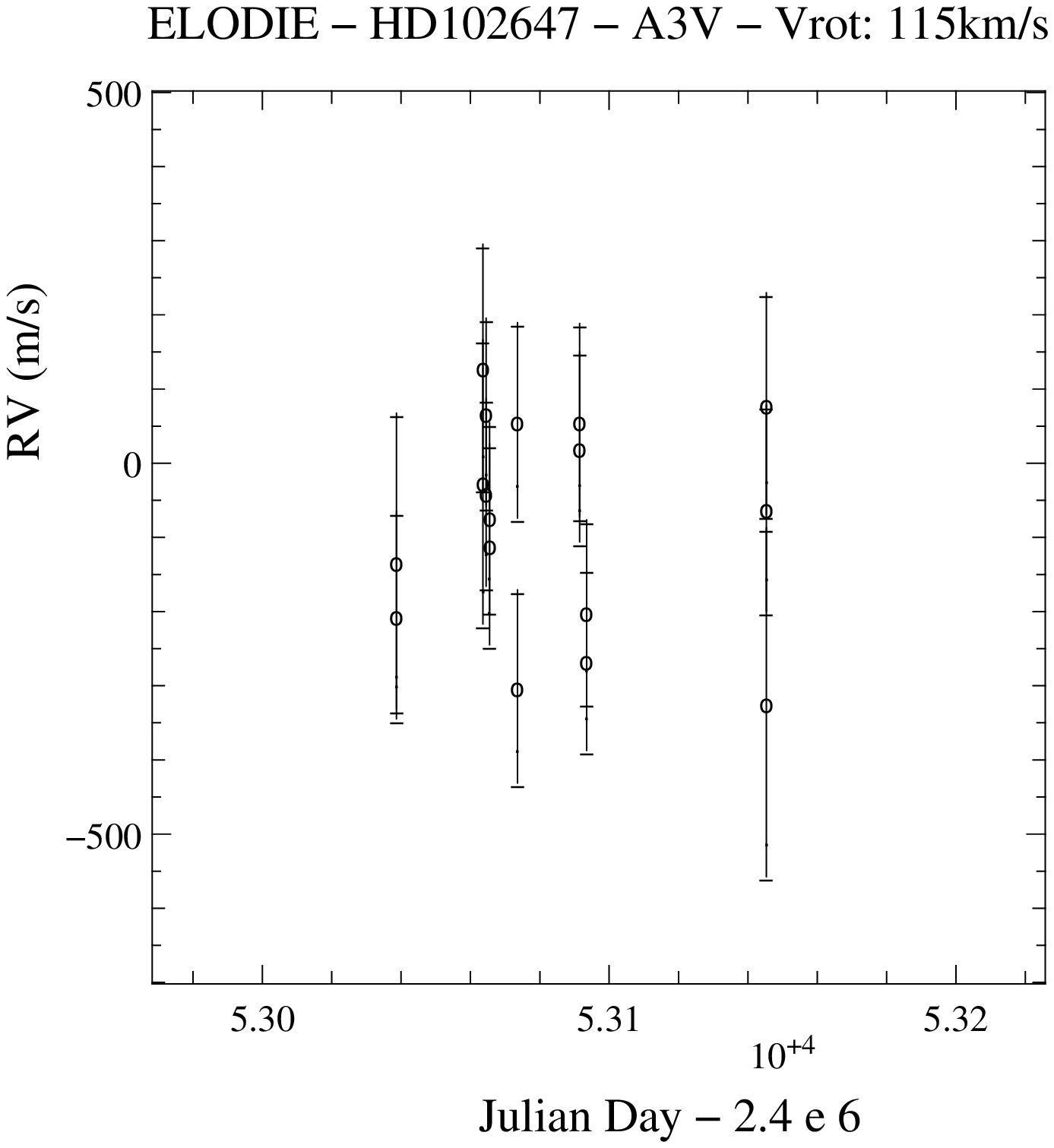} 
    \includegraphics[width=0.32\hsize]{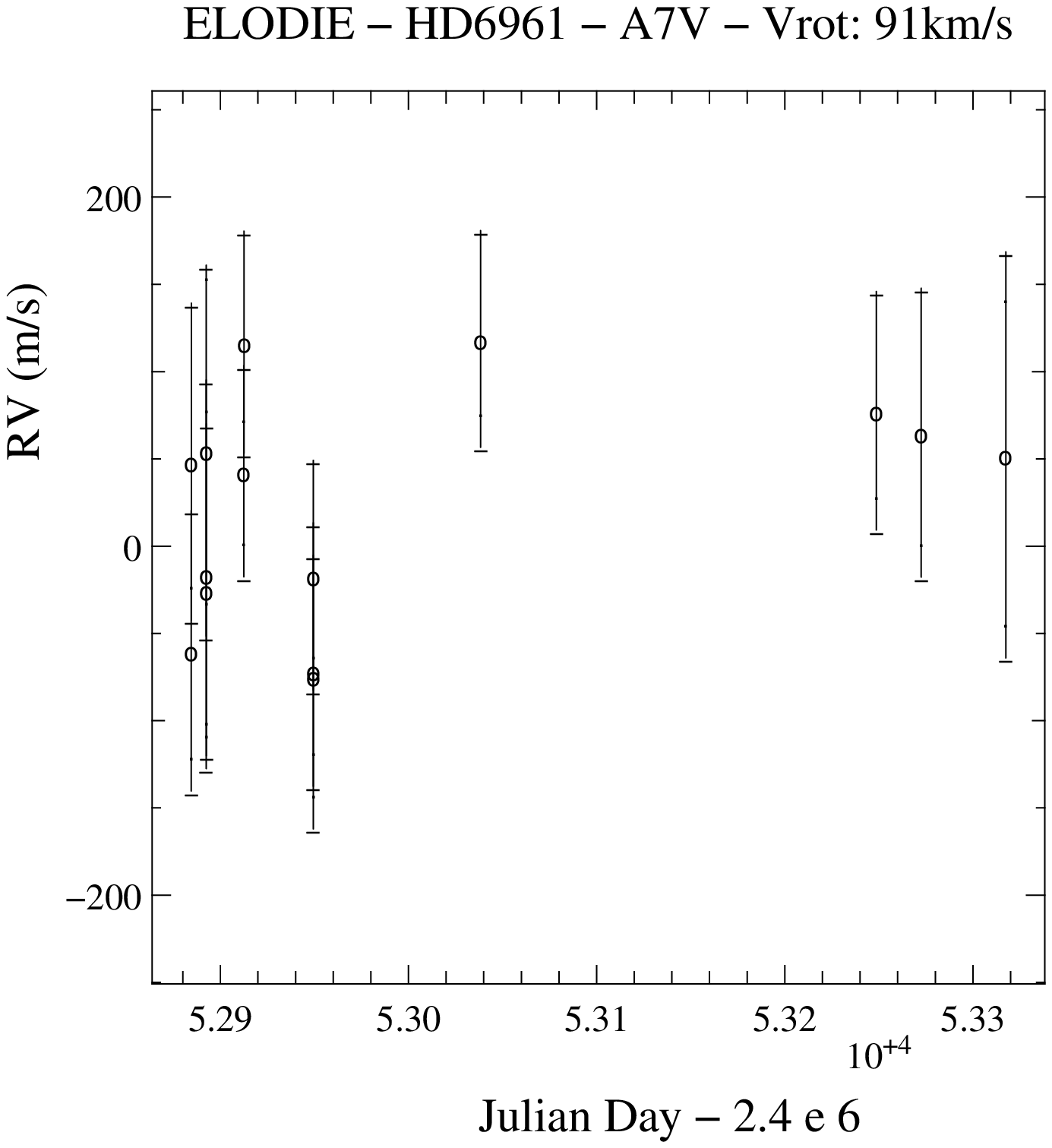}

    \vspace{0.2cm}

    \includegraphics[width=0.32\hsize]{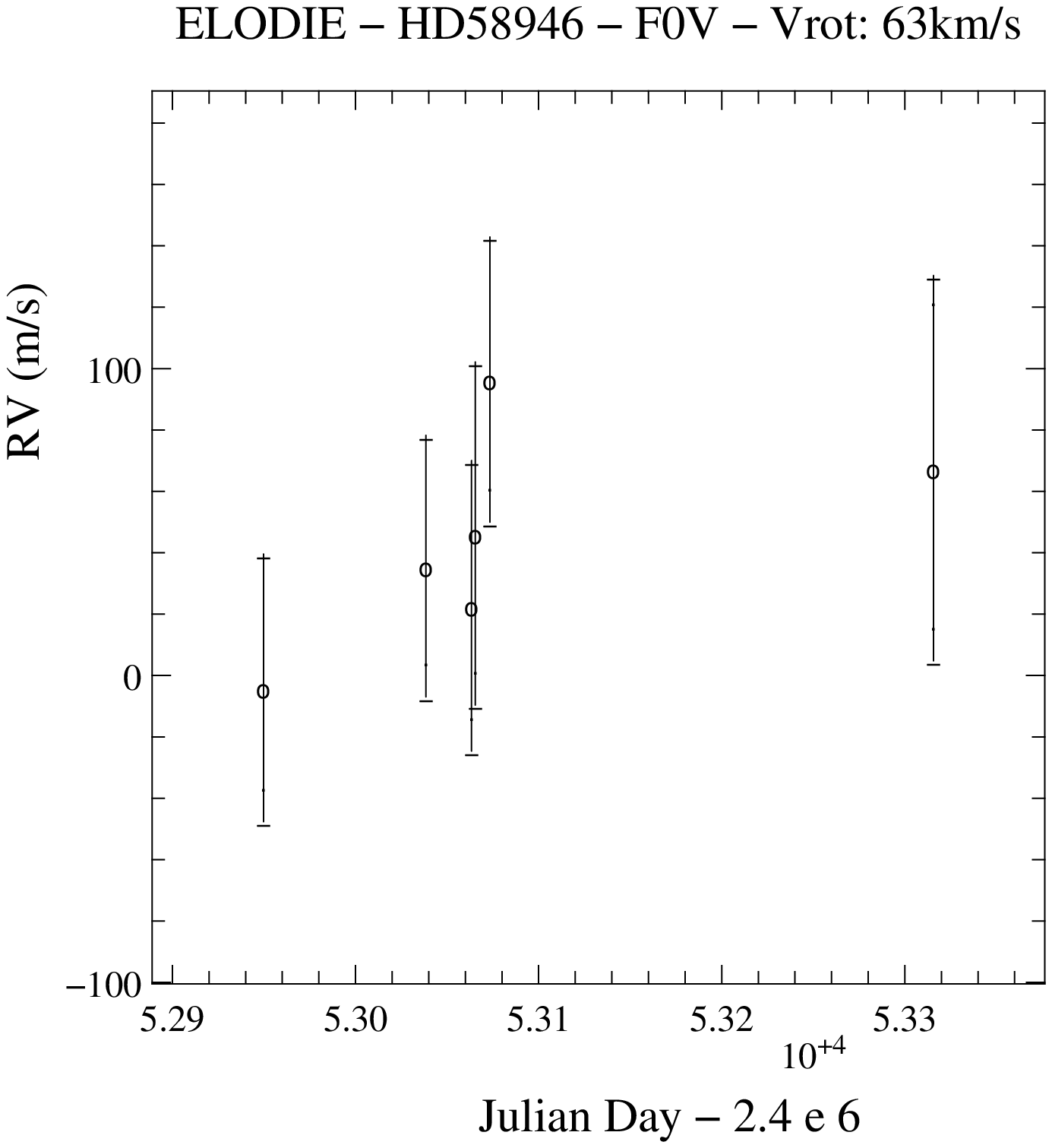} 
    \includegraphics[width=0.32\hsize]{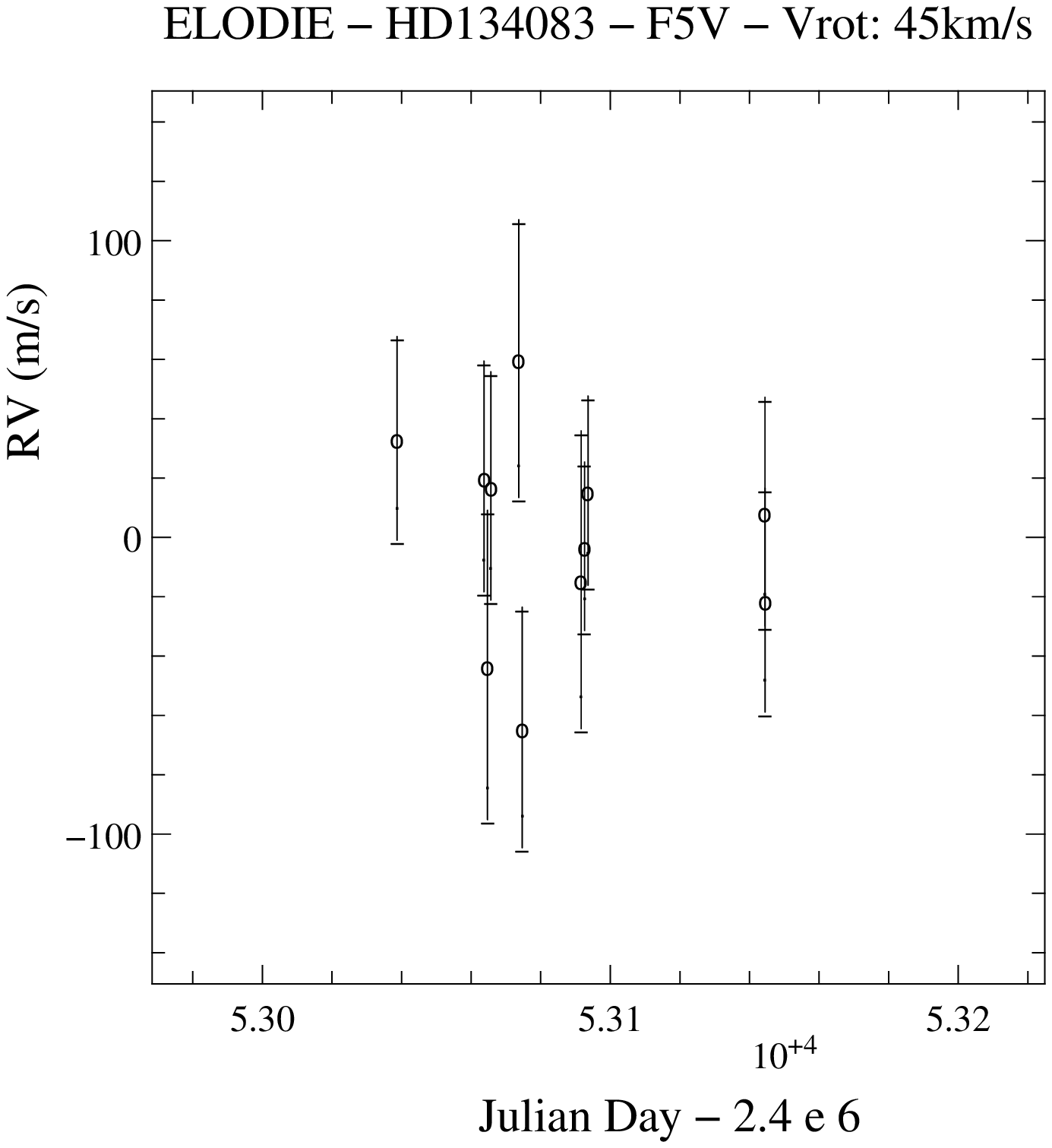} 
    \includegraphics[width=0.32\hsize]{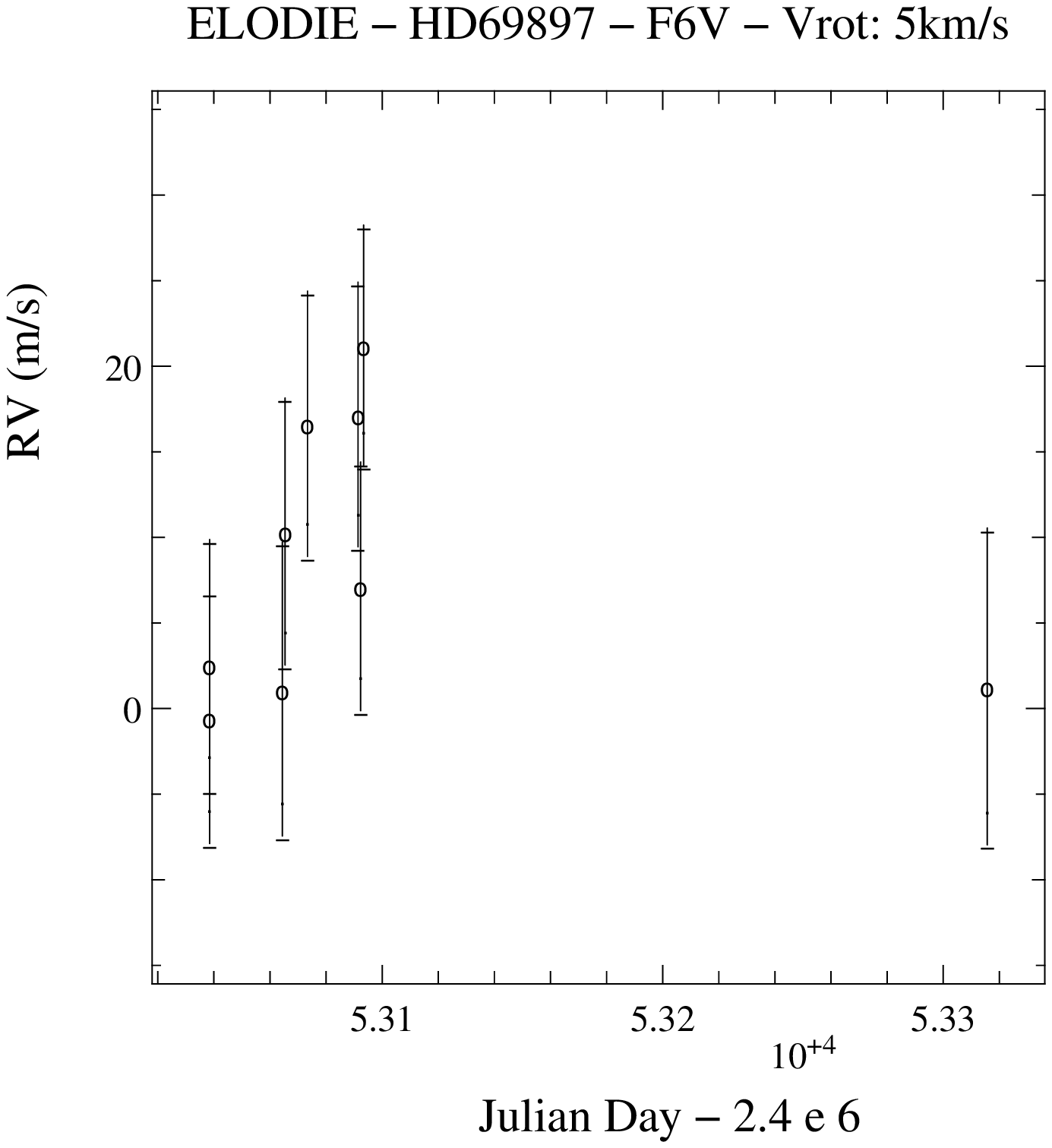}

    \caption{Examples of radial velocity measurements obtained in the
    case of stars constant in radial velocity, given the present error
    bars.}
    \label{t_E_VR}
  \end{figure*}

 \begin{figure}[t!]
   \centering
    \includegraphics[height=0.415\hsize,width=0.43\hsize]{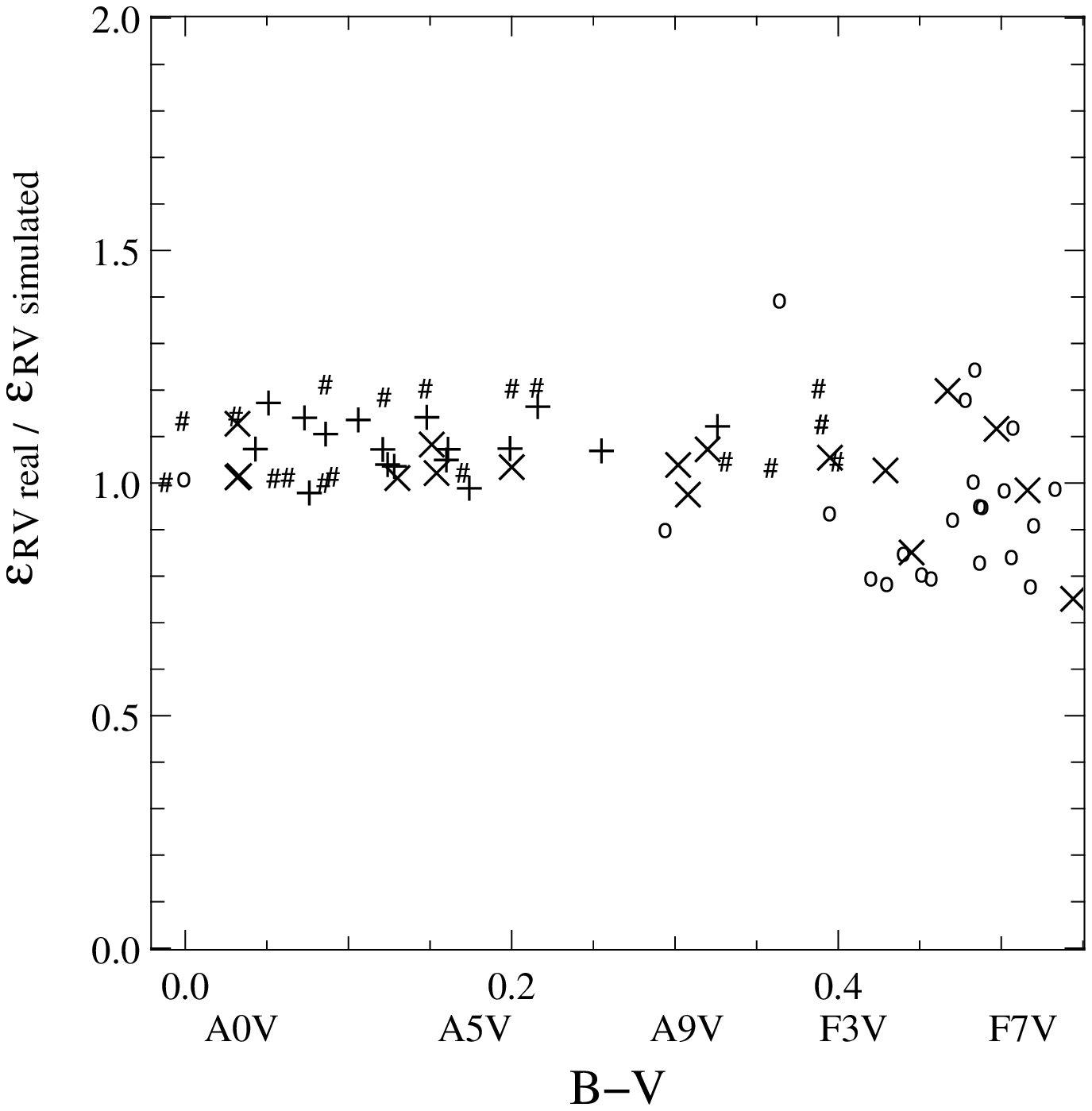}
    \includegraphics[bb=96 271 474 671,height=0.415\hsize,width=0.43\hsize,clip]{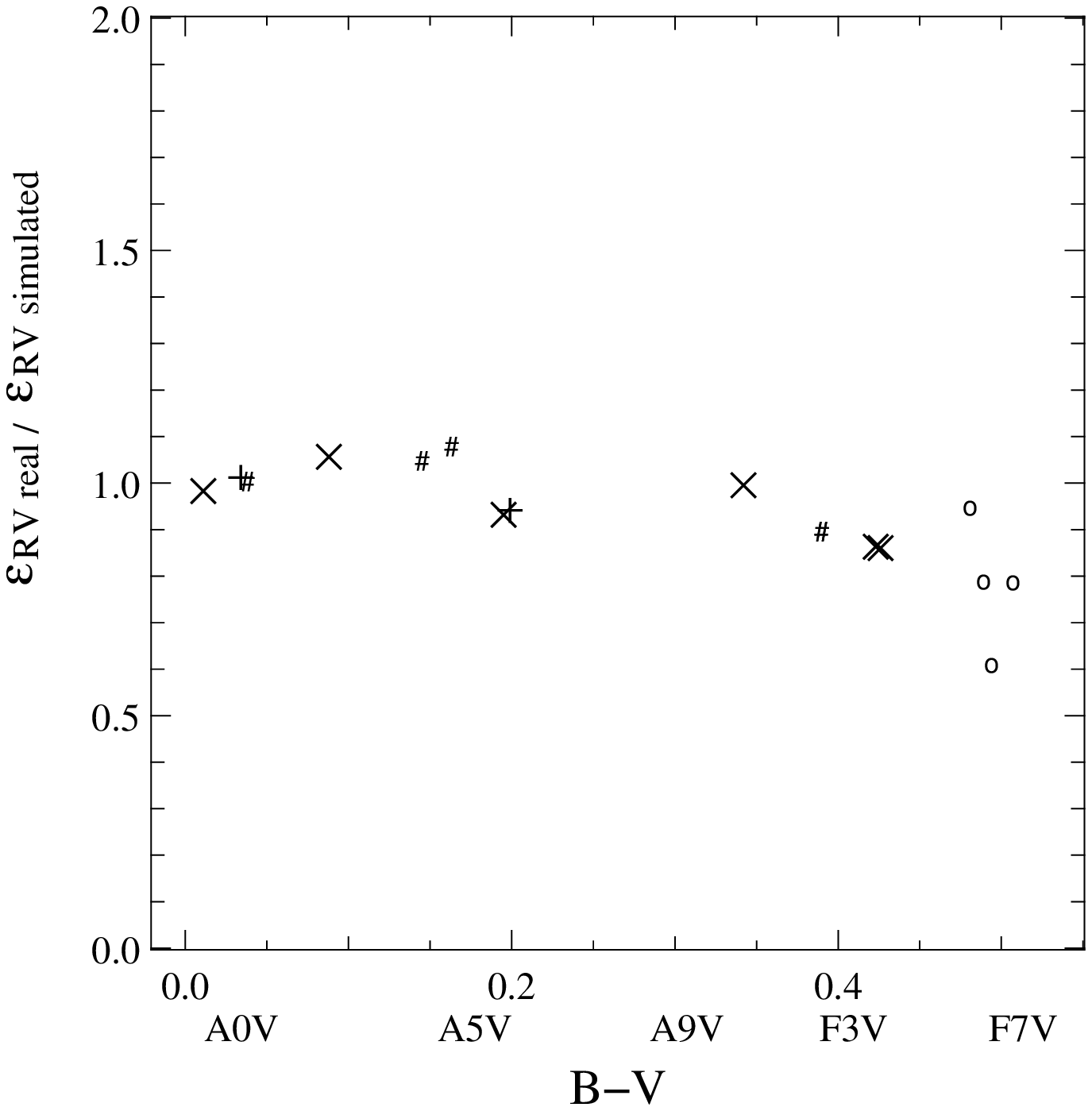}
    \caption{Radial velocity uncertainties obtained in real cases
      ($\epsilon_{\mathrm{RV real}}$) divided by
      simulated radial velocity uncertainties($\epsilon_{\mathrm{RV
      simulated}}$), with {\small ELODIE} (left) and {\small HARPS} (right).
      Symbols are the same as in Fig.~\ref{t_EH_VR_ectpeps}.
    }
    \label{EH_eps_norm_rc-eps_RV}
  \end{figure}

  Radial velocity uncertainties observed and simulated
  (normalized to the same S/N)
  are in good agreement (see Fig.~\ref{EH_eps_norm_rc-eps_RV}): in
  most cases, differences are smaller than 20 \%. Uncertainties
  observed seem slightly larger than the ones simulated, except for
  stars with B-V $\geq$ 0.4 and $v\sin{i}$ $\leq$ 20
  km\,s$^{\rm -1}$: spectra may be
  different from one to the other in real cases, resulting in imperfections
  in the reference spectrum, hence larger uncertainties; these
  imperfections could be negligible for late type
  stars, given the large number and depth of the lines.

  \subsection{Stars constant in radial velocity}
  Examples of radial velocity measurements obtained with {\small ELODIE} for
  stars that appear to be constant in
  radial velocity (given our temporal spans) are displayed in
  Fig.~\ref{t_E_VR}, and the corresponding dispersions and
  uncertainties are given in Table~\ref{Table_ectp_sigv}. The radial
  velocity dispersions are consistent with the uncertainties: this
  confirms the quality of the estimation of
  the later.

  \begin{table}
    \caption{{\small ELODIE} orbital solution for HD48097 and Tau Boo.}
    \label{Table_hd48097_par}
    \begin{center}
      \begin{tabular}{l l c c}
        \hline
        Parameter     &                          & HD\,48097       &Tau Boo \\
        \hline
        P             & [days]                   & 552$\pm$17     & 3.3135$\pm$0.0014 \\
        T             & [JD-2450000]             & 3005$\pm$40    & 1.27$\pm$0.03  \\
        e             &                          & 0.10$\pm$0.03  & 0 (fixed) \\
        $\gamma$      & [km\,s$^{\rm -1}$]            & -0.16$\pm$0.23&-0.0650$\pm$0.004 \\
        $\omega$      & [deg]                    & 304$\pm$29     & 0 (fixed) \\
        K             & [km\,s$^{\rm -1}$]             & 5.67$\pm$0.11& 0.476$\pm$0.005 \\
        N$_{\rm meas}$         &                          & 15        & 7    \\
        $\sigma(O-C)$ & [km\,s$^{\rm -1}$]            & 0.19      & 0.014 \\
        \hline
        $a_1$ sin $i$ & [AU]                     & 0.286          &  1.5$10^{-4}$\\
        f(m)          & [M$_{\odot}$]            & 1.03$10^{-2}$  &  3.7$10^{-8}$\\
        $m_1$         & [M$_{\odot}$]            & 2.7            &  1.2\\
        $m_2$ sin $i$ & [M$_{\rm Jup}$]              & 430         &  3.9\\
        $a$           & [AU]                     & 1.9            &  0.046\\
        \hline
      \end{tabular}
    \end{center}
  \end{table}

  \begin{figure}[t!]
    \centering
    \includegraphics[width=0.9\hsize]{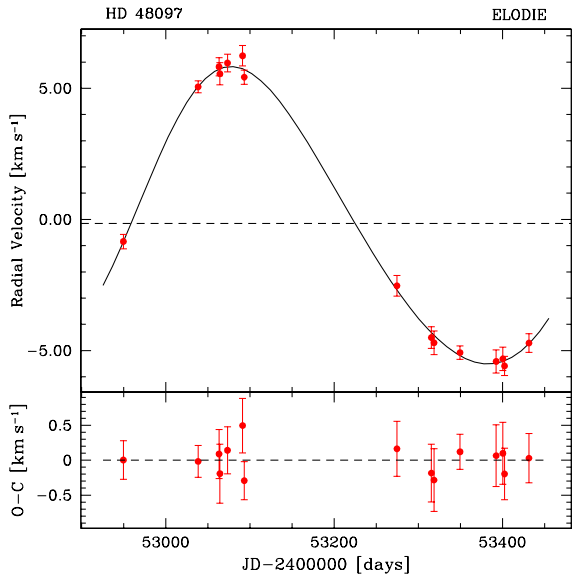}
    \caption{{\small ELODIE} radial velocity data and orbital solutions for
    HD48097. Top: Radial velocities. Bottom: Residuals.}
    \label{hd48097}
  \end{figure}

  \subsection{HD48097: a binary system}
  
  Among stars variable in our radial velocity measurements, we report here a binary
  detection in HD\,48097 (HIP\,32104), an A2V, $v\sin{i} = 90 km\,s^{\rm -1}$ star,
  with B-V = 0.063, V = 5.21, located at 43 pc from the Sun. The
  orbital parameters deduced from a
  Keplerian adjustment (Fig.~\ref{hd48097}) are displayed in Table \ref{Table_hd48097_par}. The
  companion is a star with a mimimum mass of 0.43 M$_{\odot}$, thus a
  late K or M type dwarf (and not a white dwarf for example, which should have
  evolved faster than the primary, whose mass is 2.7 M$_{\odot}$). 
  The dispersion of the residuals is 192~m\,s$^{\rm -1}$ rms, consistent with
  the radial velocity uncertainties (359~m\,s$^{\rm -1}$ on average, 283 m\,s$^{\rm -1}$ if
  normalized to $\mathrm{S/N} = 200$); the difference can
  be due to the small number of available measurements. This confirms
  the accuracy of the computed radial velocities in
  a real case.	

  \begin{figure}[t!]
    \centering
    \includegraphics[width=0.9\hsize]{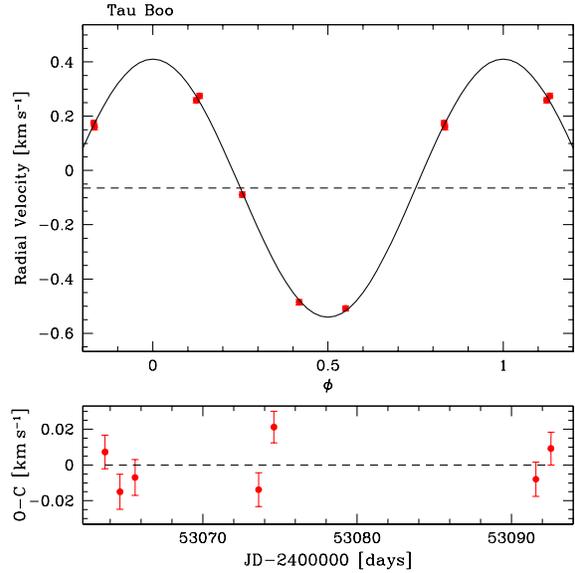}
    \caption{{\small ELODIE} radial velocity data and orbital solutions for
    Tau Boo. Top: Phased-folded velocities. Bottom: Residuals.}
    \label{hd120136}
  \end{figure}

  \subsection{HD\,120136: Tau Boo, a known planet}

  As another example, we show the measurements obtained on Tau Boo
  (HD\,120136, HR\,5185), an F7V star, with B-V = 0.48, V = 4.50, located
  15 pc from the Sun. We confirm the existence of a planet orbiting
  around this star (\cite{Butler97}).
  The orbital parameters (Fig.~\ref{hd120136}) deduced from a
  Keplerian adjustment are displayed in Table
  \ref{Table_hd48097_par}, fixing $e$ and $\omega$ to 0 given the small
  number of measurements. They are consistent with the values
  P = 3.3128 $\pm$ 0.0002 days, e = 0.02 $\pm$ 0.02, K$_1$ = 469 $\pm$ 5
  m\,s$^{\rm -1}$ previously found.
  Assuming a primary mass of 1.2~M$_{\odot}$, the minimum mass of the
  companion is still 3.9 M$_{\rm Jup}$ with a semimajor axis of 0.046 AU.

  This also confirms the accuracy of the computed radial velocities,
  in the case of a spectral type sufficiently late to have been already explored.
  A detection of a planet orbiting an F6V star can be found in
  Galland et al. (2005b).
  
  \begin{figure*}[t!]
    \centering
    \includegraphics[width=0.35\hsize]{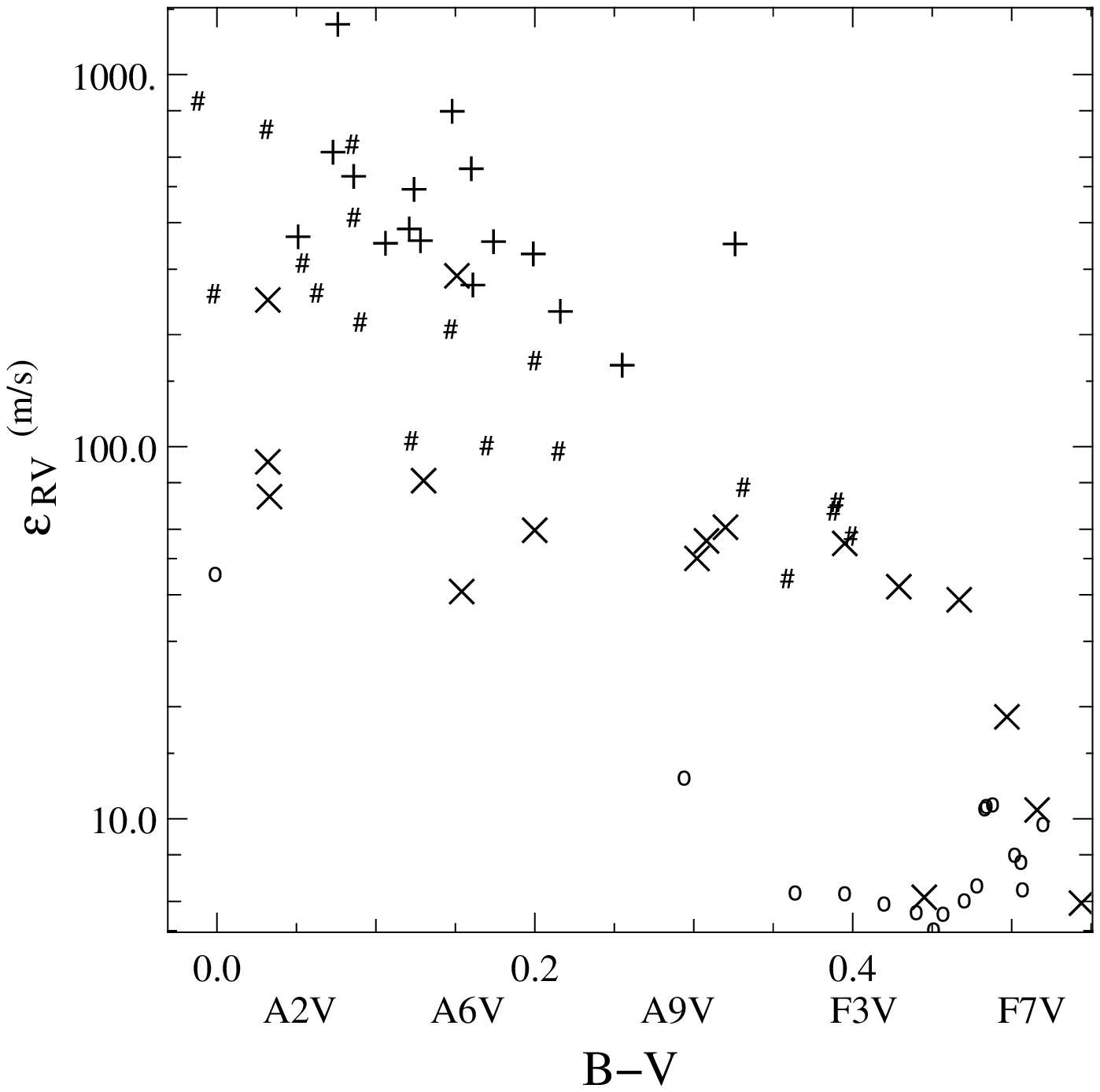}
    \includegraphics[bb=95 271 474 666,width=0.35\hsize,clip]{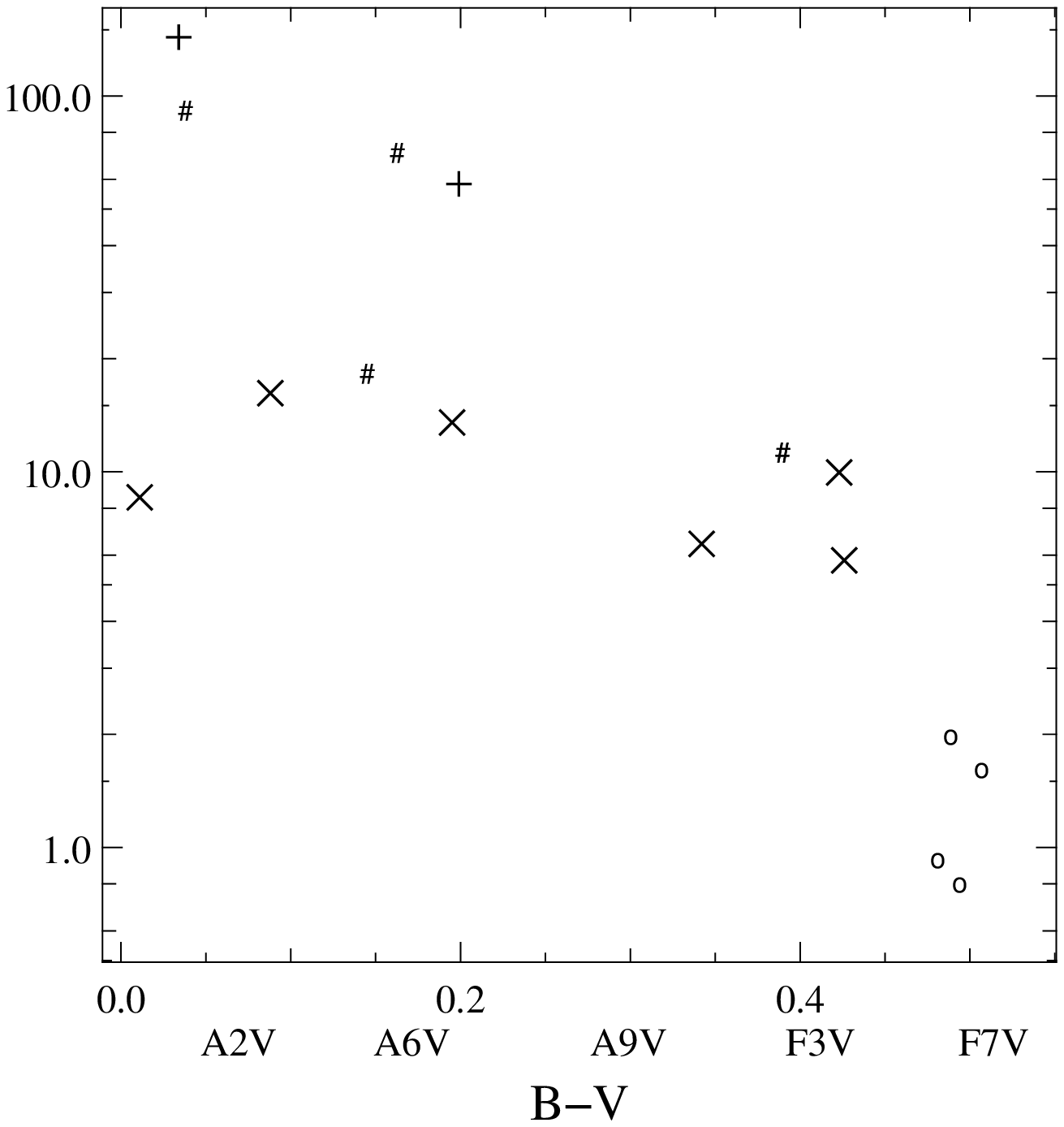} 

    \vspace{0.5cm}

    \includegraphics[width=0.35\hsize]{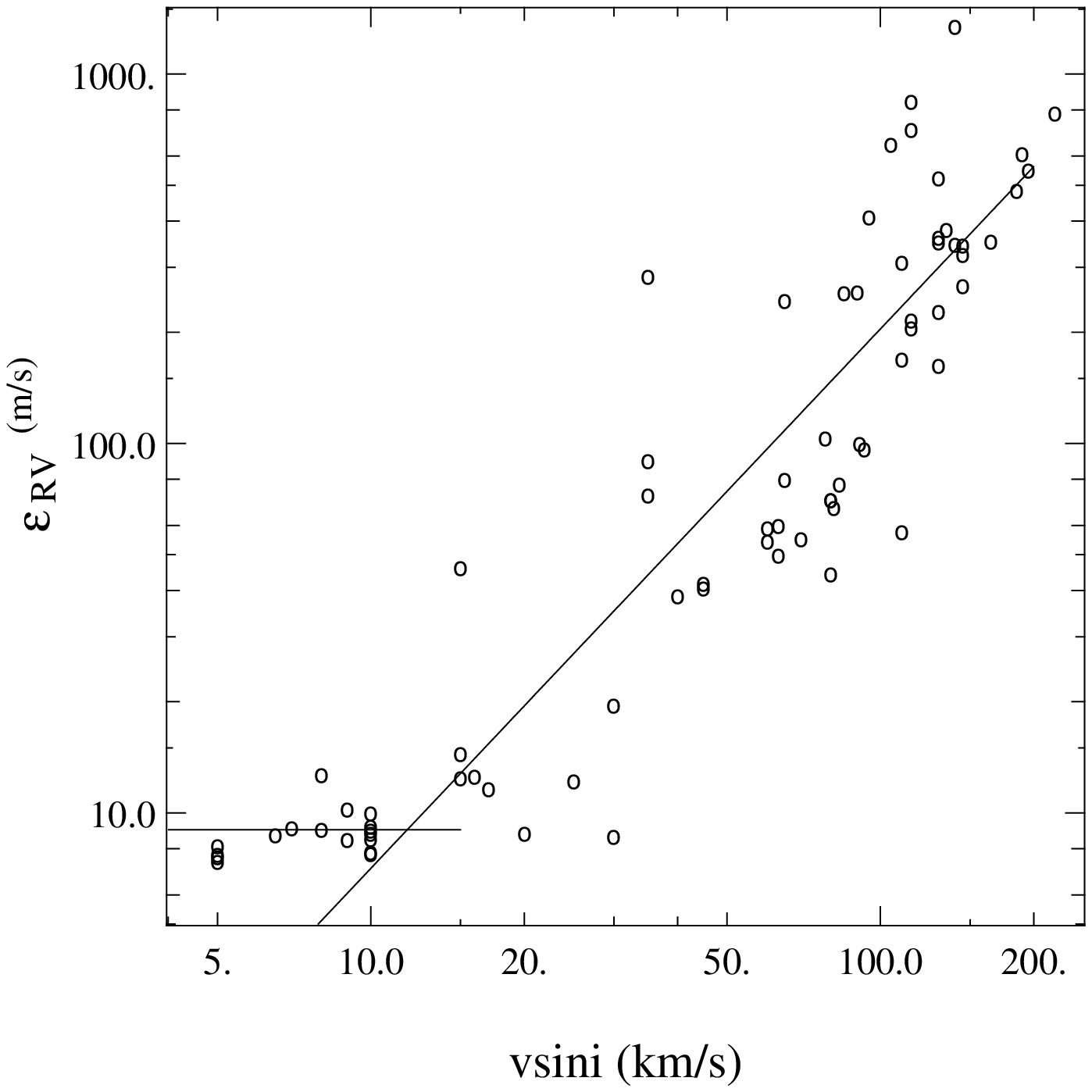}
    \includegraphics[bb=95 271 474 666,width=0.35\hsize,clip]{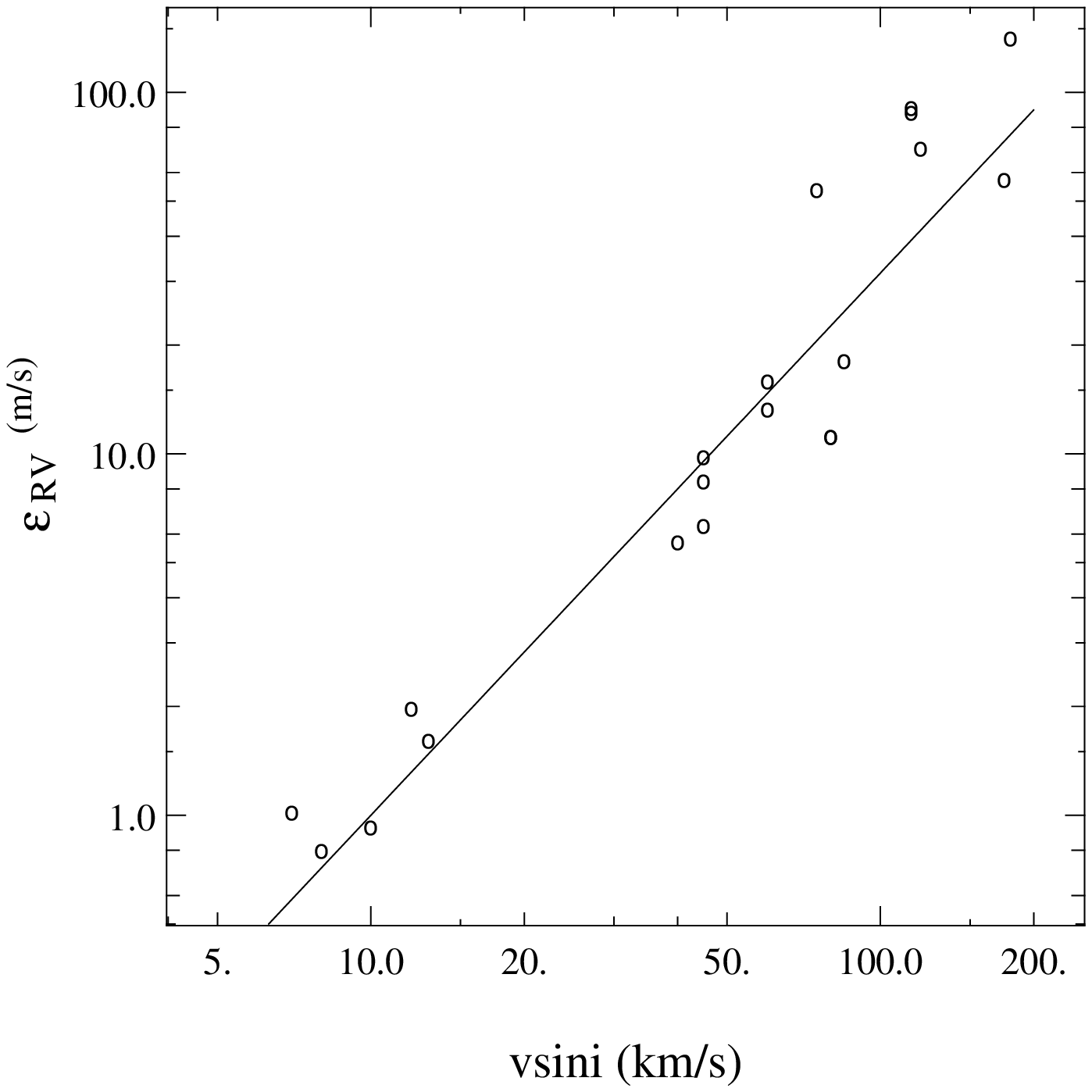} 
    \caption{Uncertainties on the radial velocities in the case of identical spectra only
    shifted in radial velocity, for all the observed stars, versus B-V
    (top; symbols are the same
    as in Fig.~\ref{t_EH_VR_ectpeps}) and $v\sin{i}$ (bottom). The
    spectra were acquired with {\small ELODIE} (left) and {\small HARPS} (right).}
    \label{t_EH_epsVR}
  \end{figure*}

  \subsection{Instability strip}
  It appears (Fig.~\ref{EH_std}) that the late A and
  early F type stars observed (B-V between 0.2 and 0.4) are often highly
  variable in radial velocity.
  This range of B-V actually corresponds to the intersection of the instability strip
  and the main sequence, where we find the pulsating $\delta$ Scuti (\cite{Handler02},
  \cite{Breger00}) and $\gamma$ Dor stars (\cite{Mathias04},
  \cite{Handler02}). These stars can be responsible for radial velocity
  variations with an amplitude up to several km\,s$^{\rm -1}$ and
  periods up to several days. 
  Additional observations are needed to check for each of the
  stars observed with B-V in the range 0.2-0.4 and with high radial velocity
  variations if they are members of the $\delta$ Scuti and $\gamma$ Dor
  groups.

  The studies on the associated photometric variations
  (\cite{Eyer97}) show that the amplitude of these variations can reach 8 mmag
  with periods of 2 days in this range of spectral types. Yet, the
  amplitude is less than 2 mmag for the other A-F type
  stars. The latter may thus be preferential targets if pulsations are not
  desired to pollute other radial velocity variations. 

  In particular, it is probable that finding planets with the
  radial velocity method will be more difficult for stars with B-V in
  the range 0.2-0.4; ways to distinguish a planetary signal from that of
  pulsations have to be investigated.

  \section{Achieved uncertainties: influence of stellar properties}

  As radial velocity uncertainties have been demonstrated to be accurately computed,
  we can discuss their values. We consider here the simulated
  radial velocity
  uncertainties, as they correspond to the case of identical spectra
  only shifted in radial velocity (keeping in mind that they are in good agreement with
  the real case, differences being less than 20~\% in most cases).

  These radial velocity uncertainties are displayed in Fig.~\ref{t_EH_epsVR}.
  We can see that they depend on the spectral type of the star (the
  later the spectral type, the smaller $\epsilon_{\mathrm{RV}}$, for a given range of $v\sin{i}$), and on
  its rotational velocity to a higher extent. A linear fit of the
  logarithm of the radial velocity uncertainty (in m\,s$^{\rm -1}$) as a
  function of the logarithm of $v\sin{i}$ (in km\,s$^{\rm -1}$) gives :

  \vspace{0.2cm}
  $\bullet$ $\epsilon_{\mathrm{RV}} = 0.16\, \hspace{0.1cm} \times v\sin{i}^{1.54} \times
  \frac{200}{S/N} $ with {\small ELODIE},
  
 \vspace{0.1cm}

  $\bullet$ $\epsilon_{\mathrm{RV}} = 0.032 \times v\sin{i}^{1.50} \times
  \frac{400}{S/N} $ with {\small HARPS}.
  \vspace{0.2cm}

  The dependance of $\epsilon_{\mathrm{RV}}$ on $v\sin{i}$ to the
  power 1.5, found with these fits, is consistent with the study of
  \cite{Bouchy01} on the fundamental photon noise limit to radial
  velocity measurements, in the case of early F type main sequence stars.
  These fits apply only in the range where lines are resolved, and if
  photon noise uncertainties are large compared to the instrumental
  (in)stability (this corresponds typically to $v\sin{i} \geq$~15~km\,s$^{\rm -1}$ with {\small ELODIE}). When
  the lines are not resolved (small  $v\sin{i}$), we reach the
  instrumental limit, represented by a horizontal
  line in Fig.~\ref{t_EH_epsVR}, in the case of {\small ELODIE}. Note that the
  dispersion around the fits results from stars with different spectral types.
  
  \begin{figure*}[t!]  
    \centering
    \includegraphics[width=0.35\hsize]{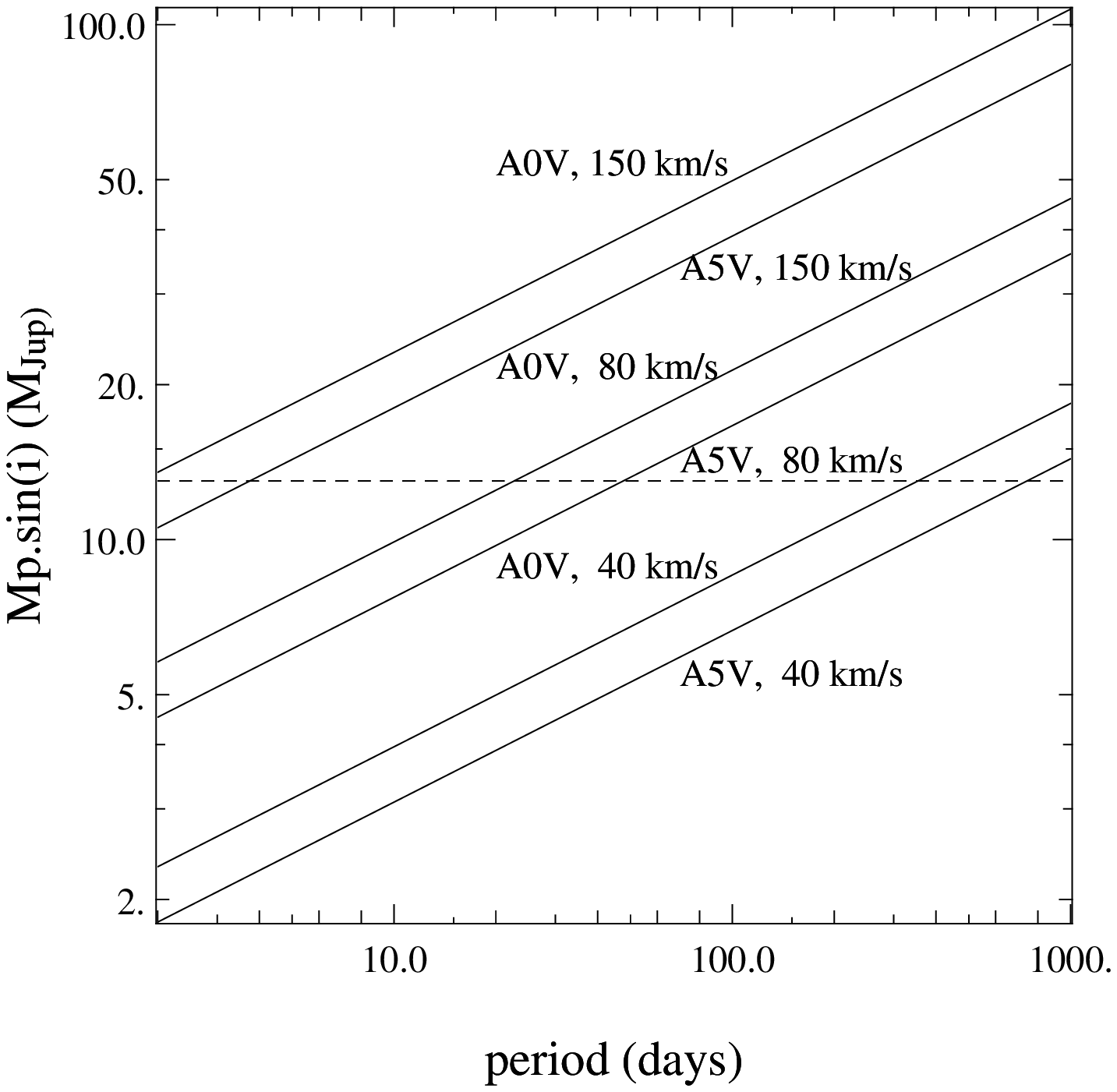}
    \includegraphics[bb= 96 271 494 666,width=0.35\hsize]{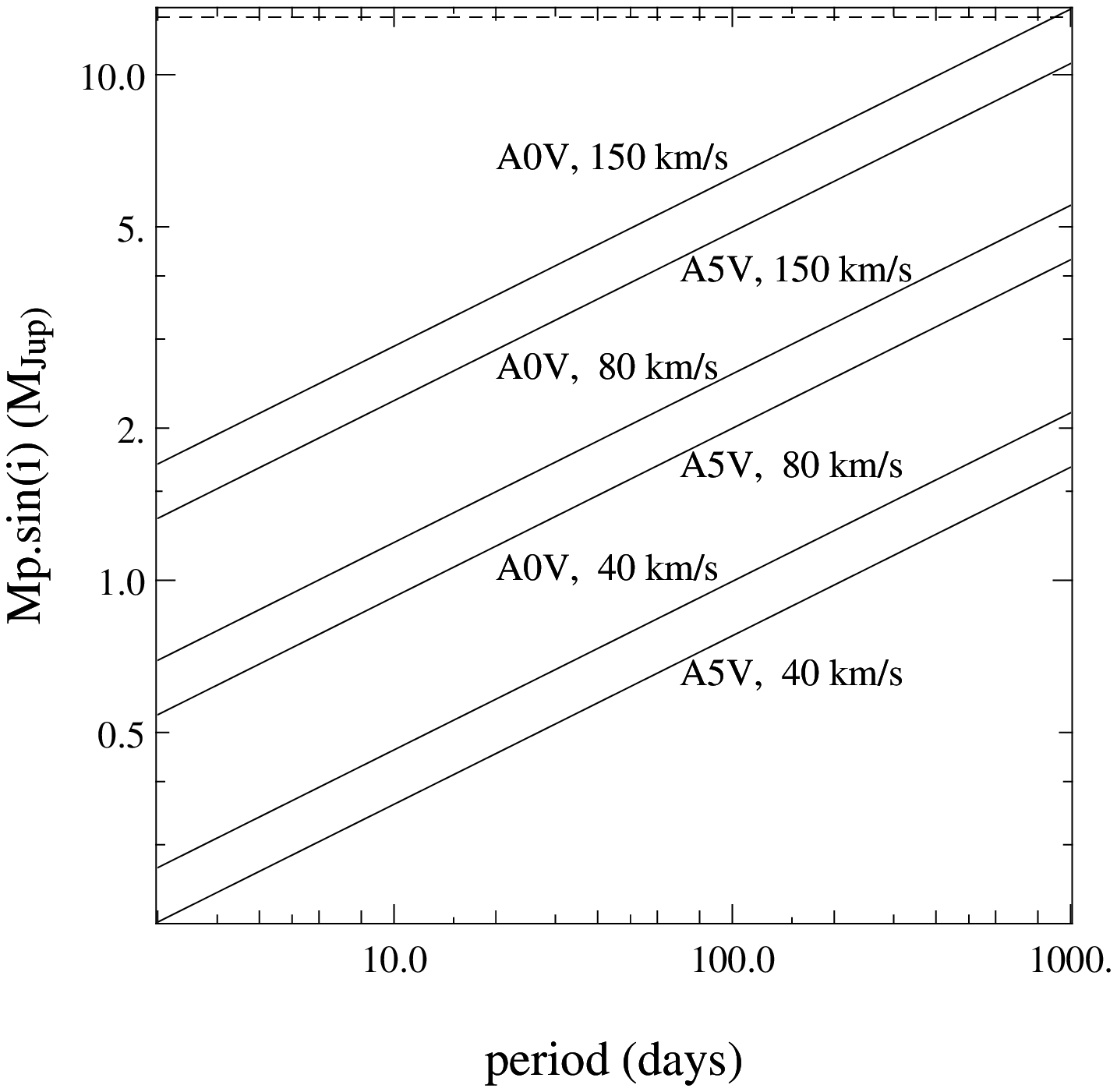}

    \vspace{0.5cm}

    \includegraphics[width=0.35\hsize]{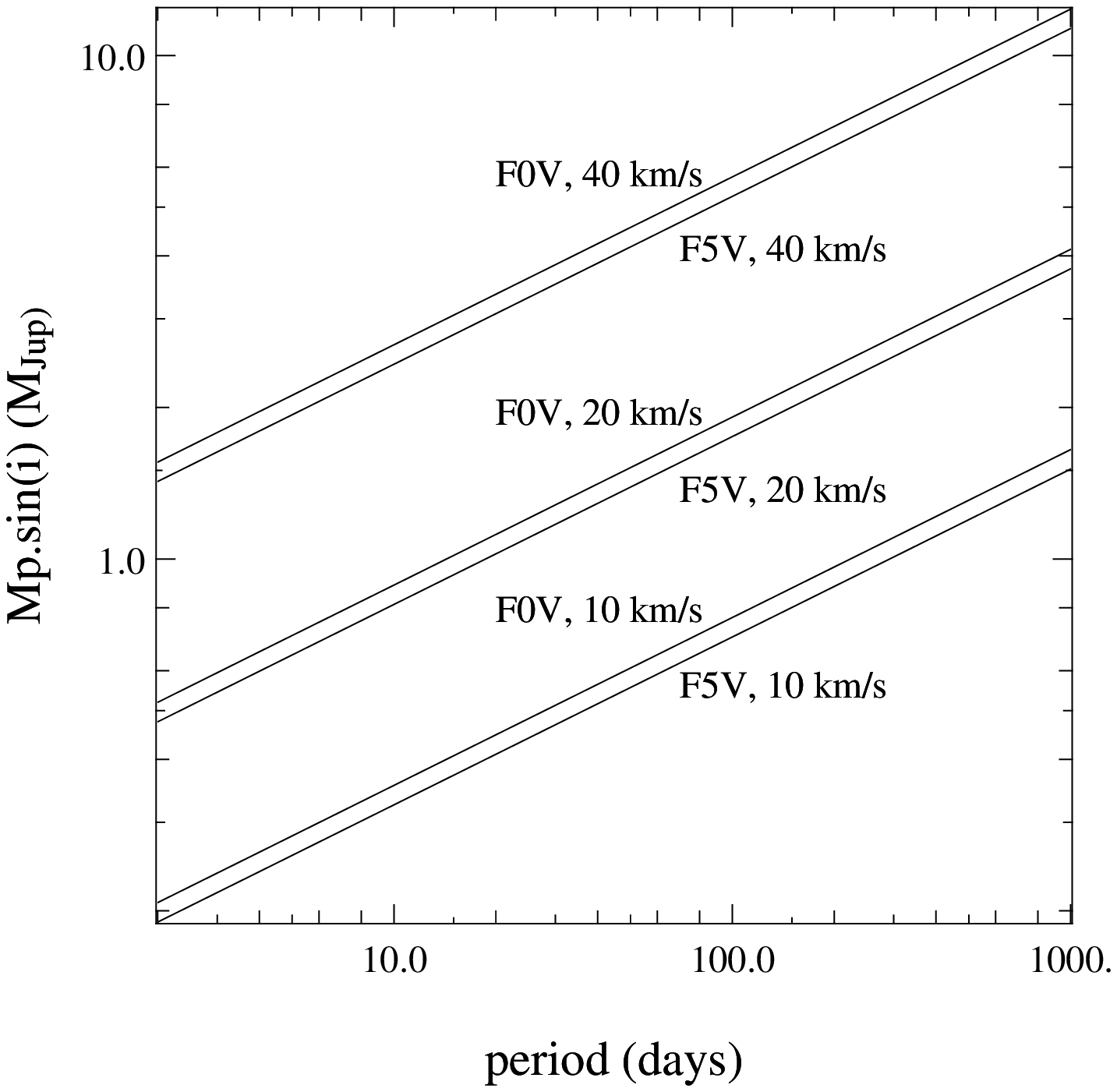}
    \includegraphics[bb= 96 271 494 666,width=0.35\hsize]{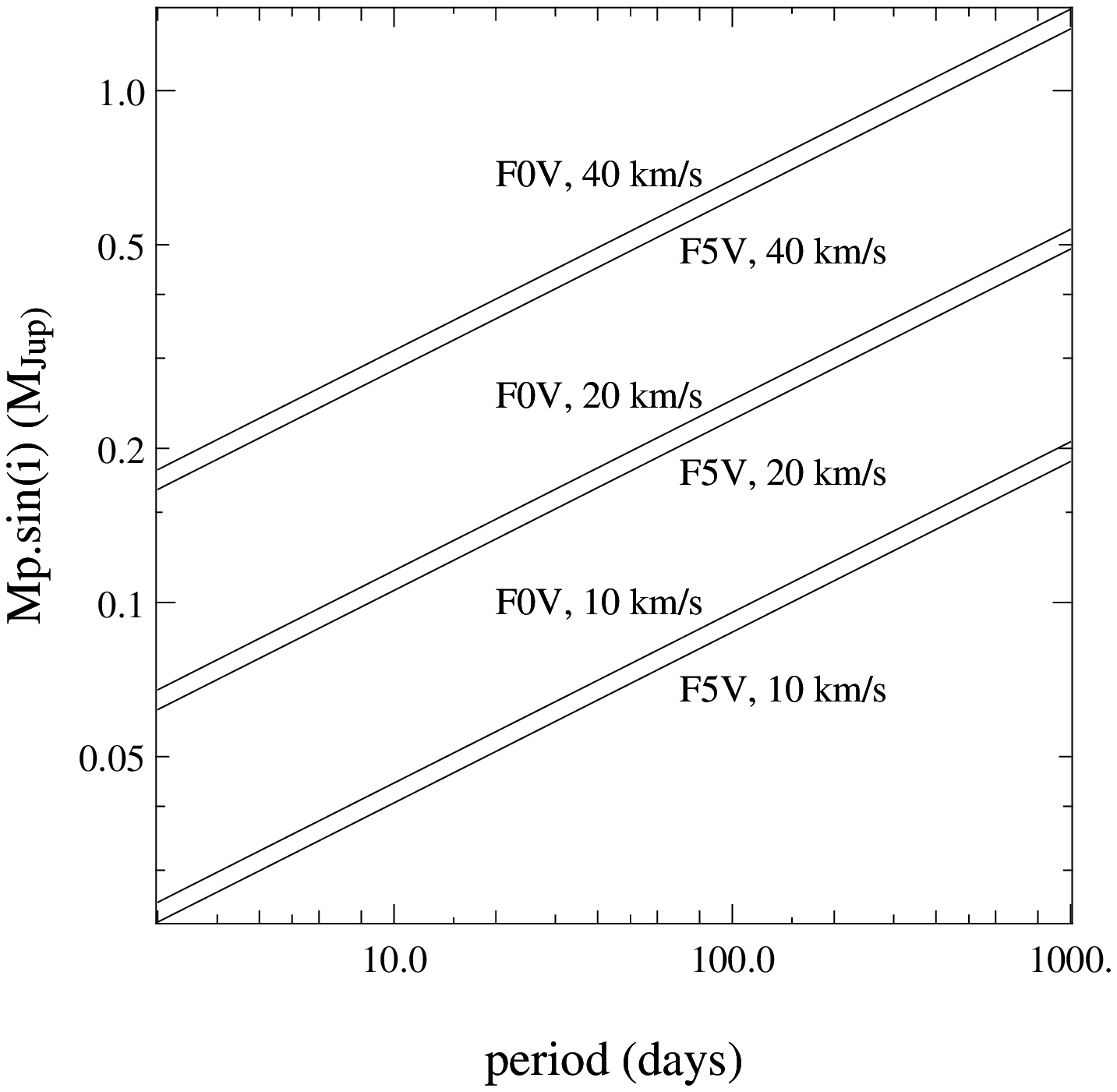}
    \caption{Mass detection limits for A (top) and F (bottom) type
    stars, using {\small ELODIE} (left) or {\small HARPS} (right).}
    \label{EH_MpT_AF}
  \end{figure*}

  The achieved uncertainties on the radial velocities appear to be a
  factor of 5$\pm$1 lower with spectra obtained with
  {\small HARPS} than with {\small ELODIE}, even for early type stars which have on
  average a large $v\sin{i}$.
  A factor of 2 comes from the S/N per pixel corresponding to these
  radial velocity
  uncertainties (200 with {\small ELODIE} versus 400 with {\small HARPS}).
  Another factor of 2 comes from an increased factor of 4 in
  the number of pixels per spectral element, i.e. a factor of 4 in the whole flux
  for a given S/N per pixel ({\small ELODIE} and {\small HARPS} have approximatively the same spectral range).
  For the radial velocity uncertainty, these two effects give a factor of 2
  $\times$ 2 $=$ 4, consistent with the factor 5$\pm$1 above.

  Note that we reach {\small ELODIE} and {\small HARPS} instrumental precision for slowly rotating stars.
      
  \section{Mass detection limits}

  In the framework of searches for low mass companions,
  we now estimate what kind of such companions can be found given
  these uncertainties.

  The mass detection limits are inferred for different orbital periods,
  given radial velocity variations of $\pm$ 3 $\epsilon_{\mathrm{RV}}$, assuming
  a circular orbit. They are displayed in Fig.~\ref{EH_MpT_AF}. The
  most important outcomes are:
  \begin{itemize}
  \item With {\small ELODIE}, the planetary domain can be reached for
    A type main sequence stars with
    $v\sin{i}$ up to 100~km\,s$^{\rm -1}$ and orbital periods less
    than 10 days, or with $v\sin{i}$ up to 40~km\,s$^{\rm -1}$ and
    orbital periods less than 1000~days. For late A type stars, the
    accessible range is $v\sin{i}$ up to 80~km\,s$^{\rm -1}$ and
    orbital periods up to 100~days. Planetary masses can be detected
    for all F type main sequence stars. 

    \vspace{0.2cm}

  \item With {\small HARPS}, the planetary domain is accessible for all A and F type
    stars, even with large $v\sin{i}$.
  \end{itemize}
  For example, with {\small ELODIE}, the mass detection limit of a 10~d period
  planet around an A5V star with $v\sin{i}$ equal to 60~km\,s$^{\rm -1}$ is 4~M$_{\rm Jup}$.
  Such massive close-in planets are not
  unexpected if a proto-planetary massive disk scales with the parent star mass.
  With {\small HARPS}, this detection limit decreases to 0.7~M$_{\rm Jup}$. 

  \section{Conclusions}

  We presented in this paper the performances of a radial velocity measurement method
  that we developed in the case of A-F type stars.
  Radial
  velocities and corresponding uncertainties are shown to be accurately computed
  both by simulations and using real cases (stars constant in radial
  velocity, the case of a binary detection and the confirmation of a
  known planet orbiting Tau Boo). 

  With regard to stellar properties, the achieved
  uncertainties $\epsilon_{\mathrm{RV}}$ depend
  on the spectral type, and above all on $v\sin{i}$, to the
  power~3/2: if $\epsilon_{\mathrm{RV}}$ is expressed in m\,s$^{\rm -1}$ and
  $v\sin{i}$ \, in km\,s$^{\rm -1}$, $\epsilon_{\mathrm{RV}}$ behaves
  typically (still with a dependance on the spectral type) as:

  \vspace{0.2cm}
  $\bullet$ $\epsilon_{\mathrm{RV}}=0.2\,\times v\sin{i}^{1.5}\times
  \frac{200}{S/N}$ with {\small ELODIE}, if \hspace{1.65cm}\, $v\sin{i}\geq$~15~km\,s$^{\rm -1}$;
  
 \vspace{0.1cm}

  $\bullet$ $\epsilon_{\mathrm{RV}} = 0.03 \times v\sin{i}^{1.5} \times
  \frac{400}{S/N} $ with {\small HARPS}.
  \vspace{0.2cm}

  In particular, we have demonstrated that it should be possible to
  detect extrasolar planets and
  brown dwarfs around such A-F type stars: detection limits arrive at the
  planetary domain for most of them.
  Given these results, we have
  begun a radial velocity survey to search for
  these low mass companions around a volume-limited sample
  of A-F main sequence stars, with {\small ELODIE} and {\small HARPS}.

  \begin{acknowledgements}
  We acknowledge support from the French CNRS. We are grateful to the
  Observatoire de Haute Provence, the Programme National de
  Plan\'etologie (INSU), and ESO for the time
  allocation, and to their technical staff.
  These results have made use of the SIMBAD database, operated at
  CDS, Strasbourg, France.    
  \end{acknowledgements}


\end{document}